\documentclass[journal]{IEEEtran}
\pdfoutput=1

\usepackage{tikz} 
\newcommand\synt[1]{\textsf{#1}}
\usepackage{cite}

\usepackage{color}
\usepackage{colortbl}
\usepackage{multirow}
\usepackage{nicefrac}
\usepackage{stmaryrd}
\usepackage{amsfonts}

\ifCLASSOPTIONcompsoc
  \usepackage[caption=false,font=normalsize,labelfont=sf,textfont=sf]{subfig}
\else
  \usepackage[caption=false,font=footnotesize]{subfig}
\fi

\usepackage{graphbox}

\usepackage[inline]{enumitem}
\usepackage{mathtools}
\usepackage{graphicx}
\usepackage{booktabs}
\usepackage[most]{tcolorbox}
\usepackage{fancybox}
\usepackage{makecell}
\usepackage[inline]{enumitem}
\usepackage[printwatermark]{xwatermark}
\usepackage{multirow}
\usepackage[binary-units,per-mode=symbol,detect-weight=true, detect-family=true,inter-unit-product =\cdot]{siunitx}

\newcommand\resq[1]{
\noindent 
\fcolorbox{green!40!black}{green!5}{\noindent 
 \parbox{0.98\columnwidth}{\noindent  #1}}\\
}

\definecolor{lightgray}{gray}{0.75}

\newtheorem{definition}{Definition}

\newcommand\timedomain{\ensuremath{\mathbb{T}}}
\newcommand\real{\ensuremath{\mathbb{R}}}

\newcommand\inputs{\ensuremath{\overline{\texttt{u}}}}
\newcommand\inputsignal{\ensuremath{u}}

\newcommand\outputs{\ensuremath{\overline{\texttt{y}}}}
\newcommand\outputsignal{\ensuremath{y}}

\usepackage[switch]{lineno}

\definecolor{myBlue}{RGB}{0,133,255}
\definecolor{myOrange}{RGB}{255,115,0}
\definecolor{myGreen}{RGB}{0,115,0}

\newcommand\phase[1]{\tikz[baseline=(X.base)]\node [draw=myBlue,fill=myBlue,thick,rectangle,inner sep=2pt, rounded corners=2pt](X){\color{white}\textbf{#1}};}

\usepackage{algorithm}
\usepackage{algpseudocode}

\begin{document}

\title{Combining Genetic Programming and Model Checking to Generate Environment Assumptions}

\author{

Khouloud~Gaaloul, Claudio~Menghi, Shiva~Nejati, Lionel~C. Briand, Yago~Isasi~Parache
\IEEEcompsocitemizethanks{
This work has been submitted to the IEEE for possible publication. Copyright may be transferred without notice, after which this version may no longer be accessible.\protect\\
\protect\\
\IEEEcompsocthanksitem K.~Gaaloul and C. Menghi  are with University of Luxembourg, Luxembourg.\protect\\
E-mail: \{khouloud.gaaloul,claudio.menghi\}@uni.lu
\IEEEcompsocthanksitem S.~Nejati and L.~Briand are with University of Ottawa, Canada and University of Luxembourg, Luxembourg.\protect\\
E-mail: \{snejati,lbriand\}@uottawa.ca.
\IEEEcompsocthanksitem Y.~I. Parache is with LuxSpace, Luxembourg.\protect\\ E-mail: isasi@luxspace.lu

}

}

\IEEEtitleabstractindextext{
\begin{abstract}
Software verification may yield spurious failures when environment assumptions are not accounted for. Environment assumptions are the expectations that a system or a component makes about its operational environment and are often specified in terms of conditions over the inputs of that system or  component. In this article, we propose an approach to automatically infer environment assumptions for Cyber-Physical Systems (CPS). Our approach improves the state-of-the-art in three different ways: First, we learn assumptions for complex CPS models involving signal and numeric variables; second, the learned assumptions include arithmetic expressions defined over multiple variables; third, we identify the trade-off between soundness and informativeness of environment assumptions and demonstrate the flexibility of our approach in prioritizing either of these criteria. 

We evaluate our approach using a public domain benchmark of CPS models from Lockheed Martin and a component of a satellite control system from LuxSpace, a satellite system provider. The results show that our approach outperforms state-of-the-art techniques on learning assumptions for CPS models, and further, when applied to our industrial CPS model, our approach is able to learn assumptions  that are sufficiently close to the assumptions manually developed by engineers to be of practical value.

\end{abstract}

}

\maketitle
\IEEEdisplaynontitleabstractindextext
\IEEEpeerreviewmaketitle    

\section{Introduction}
\label{sec:intro}
\IEEEPARstart{S}{oftware} verification can be applied either at component-level or system-level; and be performed  exhaustively using formal methods or partially via automated testing.  Irrespective of the level at which verification is conducted or the technique used for verification,  the results of verification may include spurious failures due to the violation of implicit or environment assumptions. Environment assumptions are 
the expectations that a system or a component makes about its operational environment and are often specified in terms of conditions over the inputs of that system or component. 
Any system or  component is expected to operate correctly in its operational environment, which includes its physical environment and other systems and components interacting with it. 
However, attempting to verify systems or  components for a more general environment than their expected operational environment may lead to overly pessimistic results or to detecting spurious failures~\cite{giannakopoulou2002assumption}.

Environment assumptions are rarely fully documented for software systems~\cite{10.1145/3270112.3270115}. Manual identification of  environment  assumptions  is tedious and time-consuming. This problem is exacerbated for cyber-physical systems (CPS) that often have complex,  mathematical behavior. For CPS, we often need to apply verification at the component-level, e.g., when exhaustive and formal verification do not scale to the entire system. For most practical cases, engineers simply may not have sufficient information about the details of every component of the CPS under analysis and are not able to develop sufficiently detailed, useful and accurate environment assumptions for individual CPS components.

The problem of synthesizing environment assumptions has been extensively studied in the area of formal verification and compositional reasoning (e.g.,~\cite{derler2013cyber,SANGIOVANNIVINCENTELLI2012217,10.1007/BFb0028765,10.1007/3-540-45449-7_11,10.1145/3372020.3391557}). There have
been approaches to automate the generation of environment assumptions  in the context of assume-guarantee
 reasoning using an exact learning algorithm for regular languages and finite state automata~\cite{giannakopoulou2002assumption,giannakopoulou2004assume,DBLP:conf/tacas/CobleighGP03}. These approaches, however, assume that the components under analysis and their environment can be specified as abstract finite-state machines. Such state machines are not expressive enough to capture
 CPS, and in particular quantitative and numerical CPS components and their continuous behavior.
  Besides,
  CPS components may not be readily specified in (abstract) state machine notations and specifying them into this notation may require considerable additional effort, which may not be feasible or beneficial.

In our earlier work, we proposed EPIcuRus~\cite{epicurus20} (assumPtIon geneRation approach for CPS) to automatically generate assumptions for CPS models.  EPIcuRus addresses the limitations discussed above by generating assumptions for CPS models specified in Simulink\textsuperscript{\tiny\textregistered}, which is a dynamic modeling language commonly used for CPS development  and can specify complex mathematical and continuous functions over numeric and signal variables. 
EPIcuRus receives as input a CPS  component $M$ specified in Simulink\textsuperscript{\tiny\textregistered} and a requirement $\phi$. 
It automatically infers a set of environment assumptions (i.e., conditions) on the inputs of $M$ such that $M$ satisfies $\phi$ when its inputs are restricted by those conditions. EPIcuRus uses search-based testing  to generate a set of test cases for $M$ exercising requirement $\phi$ such that some test cases are passing and others are failing. The generated test cases and their pass/fail results are then fed into a decision tree classification algorithm~\cite{fitctree}  to automatically infer an assumption $\texttt{A}$ on the inputs of $M$ such that $M$ is likely to satisfy $\phi$ when its inputs are restricted by $\texttt{A}$. Model checking is then used to validate the soundness of \texttt{A}. Specifically, model checking  determines if $M$ provably satisfies  $\phi$ when its inputs are constrained by $\texttt{A}$. If so,  $\texttt{A}$ is a  sound assumption. Otherwise, EPIcuRus continues until either a sound assumption is found or the search time budget runs out.

While EPIcuRus was effective in computing sound assumptions involving signal and numeric variables for industrial Simulink\textsuperscript{\tiny\textregistered} models,  the structure of the  assumptions generated by EPIcuRus was  rather simple. Specifically, EPIcuRus could only learn conjunctions of conditions where each condition compares exactly one signal or numeric variable with a constant using a relational operator.  This is because EPIcuRus uses decision tree classifiers that can only infer such simple conditions. In our experience, however, assumptions produced by EPIcuRus, while being sound, are not the most informative assumptions that can be learned for many CPS Simulink\textsuperscript{\tiny\textregistered} models. In this paper, the goal is to learn an assumption that is not only sound (i.e., makes the component satisfy its requirements), but is also informative (i.e., is ideally the weakest  assumption or among the weaker assumptions that make the component satisfy the  requirement under analysis). For example, the actual assumption of our industrial case study---the attitude control component of the ESAIL maritime micro-satellite---is in the following form: $\texttt{A}_1 \Coloneqq \forall t \in [0,1]: \alpha \cdot x(t) + \beta \cdot y(t) < c$, where $x$ and $y$ are signals defined over the time domain $[0,1]$.  But EPIcuRus, when relying on decision trees, is not able to learn any assumption in that form and instead learns assumptions in the following form: $\texttt{A}_2 \Coloneqq \forall t \in [0,1]: \alpha' \cdot x(t) < c' \wedge \beta' \cdot y(t) < c'' $. Assumptions in the latter form, even though sound, are less informative than the actual assumptions.

In this paper, we extend EPIcuRus to learn assumptions  containing conditions that relate multiple signals by both arithmetic and relational operators. We do so using genetic programming (GP)~\cite{koza1992genetic,poli2008field,banzhaf1998genetic,GPMatlab,madar2005genetic}. Provided with a grammar for the assumptions that we want to learn, GP is able to generate assumptions that structurally conform to the  grammar~\cite{montana1995strongly}, and in addition, maximize objectives that increase the likelihood of the  soundness and informativeness of  the generated assumptions. EPIcuRus still applies model checking to the assumptions learned by GP to conclusively verify their soundness. The informativeness, however, is achieved through GP and partly depends on the structural complexity of the learned assumptions. Any assumption that can be structurally generated by our grammar is in the search space of GP. Therefore,  EPIcuRus with GP has more flexibility compared to the old version of EPIcuRus and can search through a wider range of structurally different assumption formulas to build more expressive assumptions that are likely to be more informative as well.

Note that, in the context of CPS, as assumptions become more informative and structurally complex, establishing their soundness becomes more difficult as well. As discussed above, soundness can only be established via exhaustive verification (e.g., model checking).  In our experience with industrial CPS Simulink\textsuperscript{\tiny\textregistered} models, model checkers fail to provide conclusive results by either proving or refuting a property when the assumption used to constrain the model inputs becomes structurally complex (e.g., when it involves arithmetic expressions over multiple variables). Therefore, the more informative the assumptions, the less effective exhaustive verification tools in proving their soundness, and vice-versa. Hence, if guaranteed soundness is a priority, engineers may have to put up with less informative assumptions, and conversely, they can have highly informative assumptions whose soundness is not proven.

We evaluated EPIcuRus using two separate sets of models: First, we used a public-domain benchmark of Simulink\textsuperscript{\tiny\textregistered} models provided by Lockheed Martin~\cite{lockheedmartin}, a company working in the aerospace, defense, and security domains; second, we used a more complex model of the attitude control component of a microsatellite  provided by LuxSpace~\cite{luxspace}, a satellite system provider.
EPIcuRus successfully computed assumptions for $18$ requirements of four benchmark models from Lockheed Martin~\cite{lockheedmartin} and one requirement of the attitude control component from LuxSpace. Note that, among all of our case study models, only these requirements needed to be augmented with environment assumptions to be verified by a model checker.

We check each requirement for model inputs conforming to different signal shapes, specifying different ways the values of input signals change over time. We refer to signal shapes as input profiles and define a number of specific input profiles in our evaluation. Note that not all input profiles are valid for every model and requirement. In total, we consider $32$ combinations of requirements and  input profiles.  Our evaluation targets two questions:  if genetic programming (GP) can outperform decision trees (DT) and random search (RS) in generating informative and sound assumptions (RQ1), and
if the assumptions learned by EPIcuRus are useful in practice (RQ2).
For RQ1, we considered all the $32$ requirement and input profile combinations. 
Our results show that GP can learn a sound assumption for $31$ out of $32$ combinations of requirements and input profiles, while DT and RS can only learn assumptions for $26$ and  $22$ combinations, respectively. The assumptions computed by GP are also significantly ($20$\% and $8$\%) more informative  than those learned by DT and RS. 
For RQ2, we considered the attitude control component from LuxSpace  since this is a representative and complex example of an industrial CPS component, and more importantly, in contrast to the public-domain benchmark,  we could interact with the engineers that developed this component to evaluate how the assumptions computed by EPIcuRus compare with the assumptions they manually wrote. Our results show that,  when EPIcuRus was configured to proritize informativeness, as opposed to proving the soundness of assumptions, learned assumptions were syntactically and semantically close to those written by engineers. 
Conversely, when learning assumptions whose soundness can be verified was prioritized, EPIcuRus was able to generate sound assumptions in around six hours. Though simpler than the actual assumption, they provided useful, practical insights to engineers. We note that none of the existing techniques for learning environment assumptions is able to handle our attitude control component case study or learn assumptions that are structurally as complex as those required for this component.

\textbf{Structure.} Section~\ref{sec:running} introduces the ESAIL running example.
Section~\ref{sec:background} outlines EPIcuRus and its pre-requisites.  Section~\ref{sec:problemDef} formalizes the assumption generation problem.
Section~\ref{sec:impl} presents how  EPIcuRus is implemented. 
Section~\ref{sec:ev} evaluates  EPIcuRus, and Section~\ref{sec:disc} discusses the threats to validity.
Section~\ref{sec:rl} compares with the related work and Section~\ref{sec:conclusion} concludes the paper.

\section{ESAIL Microsatellite Case Study}
\label{sec:running}
\IEEEPARstart{O}ur case study system, the ESAIL maritime micro-satellite, is developed by 
LuxSpace~\cite{luxspace},   our industrial partner, in collaboration 
with ESA~\cite{esa} and ExactEarth~\cite{exactEarth}.
ESAIL aims at enhancing the next generation of space‐based services for the maritime sector. 
During the design phase of the satellite (i.e., development phases B-C~\cite{phases}), the control logic of the ESAIL software is specified as a Simulink\textsuperscript{\tiny\textregistered}~\cite{mathworks} model.
Before translating the Simulink\textsuperscript{\tiny\textregistered} model into code that is going to be ultimately deployed on the ESAIL satellite, 
LuxSpace engineers need to ensure that the model satisfies its requirements.

The ESAIL Simulink\textsuperscript{\tiny\textregistered} model is a large, complex and compute-intensive model~\cite{menghi2019approximationrefinement}. It contains $115$ components (Simulink\textsuperscript{\tiny\textregistered} Subsystems~\cite{subsystem}) and a large number ($2817$) of Simulink\textsuperscript{\tiny\textregistered} blocks of different types such as S-function blocks~\cite{sfunction} containing Matlab code, and some MEX functions~\cite{mex} executing  C/C++ programs  containing the behavior of external third party software components.  Due to the above characteristics,  exhaustive verification of the ESAIL Simulink\textsuperscript{\tiny\textregistered} model  (e.g., using model checking) is infeasible. For example, QVtrace~\cite{QVtrace}, an industrial model checker developed by QRA~Corp~\cite{qracorp} and dedicated to model checking Simulink\textsuperscript{\tiny\textregistered} models, cannot load the model of ESAIL due to the presence of many components the QVtrace model checker can not handle.

Even though  the entire ESAIL Simulink\textsuperscript{\tiny\textregistered} model cannot be verified using exhaustive verification, it is still desirable to identify critical components of ESAIL that are amenable to model checking. 
For example, the Attitude Control~(\texttt{AC}) component  of the attitude determination and control system (\texttt{ADCS}) of ESAIL monitors the environment in which the satellite is deployed and sends commands to its actuators according to classical control laws used for the implementation of satellites~\cite{wie1998space}. Specifically, it receives from the attitude determination system the estimated values of the speed, the attitude, the magnetic field and the sun measurements. It also receives commands from the guidance such as the target speed and attitude of the satellite. Then, it returns the commanded torque to the reaction wheel and the magnetic torquer.
\texttt{AC} has ten inputs 
and four outputs that represent the commanded torque to be fed into the actuators.
Note that some of the inputs and outputs are vectors containing several input signals, i.e., virtual vectors~\cite{virtualVectors}.
The inputs and outputs of \texttt{AC} are summarized in Table~\ref{tab:inputs}.
\texttt{AC} contains $1142$ blocks. It can be loaded in QVtrace after replacing the $19$ S-function blocks, that cannot be processed by QVtrace, with a  set of Simulink\textsuperscript{\tiny\textregistered} blocks supported by QVtrace. This activity is time-consuming and error prone. 
Every time an S-function is replaced by a set of Simulink\textsuperscript{\tiny\textregistered} blocks, to check for a discrepancy between the behaviors of the S-function and the newly added Simulink\textsuperscript{\tiny\textregistered} blocks, a set of inputs is generated and, for each input, the outputs produced by 
the S-function and the Simulink\textsuperscript{\tiny\textregistered} blocks
is compared to check for dissimilarities. 
After all the S-function blocks are removed, the model can be loaded in QVtrace, and we can have a formal proof of correctness (or lack thereof) for \texttt{AC}, which is an important component of ESAIL.

\begin{table}[]
    \centering
     \caption{Name of the Input, number of input signals in the virtual vector~(NS), and description of the input. }
    \label{tab:inputs}
    \begin{tabular}{c c c c}
    \toprule
&    \textbf{Name} &
    \textbf{NS} &
    \textbf{Description}\\
    \midrule
  \multirow{10}{*}{\textbf{Input}}  &     \texttt{q}$_t$  & 4 &  Target attitude of the satellite.  \\
  &     \texttt{q}$_e$ & 4 & Estimated attitude of the satellite.\\
 &       $\omega_t$ & 3 & Target speed of the satellite. \\
&       $\omega_e$ & 3 & Estimated speed of the satellite. \\
 &      \texttt{B} & 3 & Measured Magnetic field.  \\
  &     \texttt{Md} & 1 & The mode of the satellite.\\
   &    \texttt{Rwh} & 4 & Angular momentum of the reaction wheel.\\
    &   \texttt{Ecl} & 1 & Whether the satellite is in eclipse.\\
     &  \texttt{SF} & 1 &  Indicates if the sun sensor is illuminated.\\ 
      &  \texttt{SM} & 3 & Sun sensor measurements.\\
    \midrule
  \multirow{4}{*}{\textbf{Output}} &   \texttt{MT}  & 4 &  magnetic dipole applied to the magnetorquers.\\
  &   \texttt{MTc}  & 4 &  Current applied to the magnetorquers.\\
 &   \texttt{RW} & 3 & Torque applied to the reaction wheel.\\
 &   \texttt{RWa} & 3 & The reaction wheel's torque acceleration.\\
    \bottomrule
    \end{tabular}
\end{table}

However, some requirements may fail to hold on \texttt{AC} when it is evaluated as an independent component, while the same requirements would hold on \texttt{AC} when it is evaluated within the larger model it is extracted from. This is because in the latter case the \texttt{AC} inputs are constrained by the values that can be generated within the larger model, which is not the case when \texttt{AC} is running independently. As a result, we need to verify whether the conditions under which \texttt{AC} works are acceptable given the input values that can be generated by its larger model. This is addressed by learning assumptions guaranteeing that \texttt{AC} satisfies its requirements.

\begin{table}
    \centering
     \caption{Example requirements for the attitude control component. }
    \label{tab:requirements}
    \begin{tabular}{l p{7.5cm}}
    \toprule
  \textbf{ID} &
    \textbf{Requirement} \\
    \midrule
  $\phi_1$ & When the norm of the attitude error quaternion is less than $0.001$, the torque commanded to the reaction wheel around the x-axis (with respect to its  body frame) shall be within the range [$-0.001$,$0.001$]\SI{}{\newton\meter}.  \\
   $\phi_2$ & The magnetic moment of each magnetorquer shall be within the range [$-15$,$15$]\SI{}{\ampere\square\meter}. \\
 $\phi_3$ &The current applied to each  magnetorquer shall be within the range [$-0.176$,$0.176$]\SI{}{\ampere}.\\
  $\phi_4$ & The acceleration of each of the
  reaction wheels shall be within the range [$-0.021$,$0.021$]\SI{}{\meter\per\second\squared}.\\
    \bottomrule
    \end{tabular}
\end{table}

For example, the Simulink\textsuperscript{\tiny\textregistered} model of the \texttt{AC}  is expected to satisfy a number of requirements. Examples of these requirements are described in Table~\ref{tab:requirements}.
The requirement $\phi_1$ ensures that \texttt{AC} does not command any torque about the x-axis of the body frame to the reaction wheel,  when the satellite is already at the desired attitude. 
Reaction wheels are used to control the attitude, i.e., the orientation of the satellite, and ensure high pointing accuracy. They generate the twist applied to the satellite around a specific axis by acting on the acceleration of the reaction wheels.
To determine whether,  or not, \texttt{AC} satisfies requirement $\phi_1$, we convert the requirement into a formal property and use QVTrace~\cite{QVtrace} to verify $\phi_1$ over \texttt{AC}. However, it turns out that the requirement $\phi_1$ does not hold on \texttt{AC}. Further, using QVTrace, we cannot show that  \texttt{AC} satisfies  $\neg \phi_1$, indicating that not all of its behaviors violate the requirement of interest. Therefore, for some inputs, \texttt{AC} violates  $\phi_1$, and  for some, it satisfies  $\phi_1$. 
Note that if the model satisfies either $\phi_1$ or $\neg \phi_1$, there is no need for generating an input assumption.

One of the reasons that the \texttt{AC} does not satisfy $\phi_1$ is that, 
to ensure that the torque applied to the reaction wheel remains within \SI{-0.001}{\newton\meter} and \SI{0.001}{\newton\meter} when the norm of the attitude error quaternion is less than $0.001$, we need to constrain the inputs of \texttt{AC} by  the following assumption $\texttt{A}_1$, which we elicited,
in collaboration with the ESAIL engineers:

{\footnotesize
\begin{align}
& \texttt{A}_1 ::=\forall t  \in [0,1]:(  \overbrace{\exp(t)  +1 \geq 0}^{\texttt{P}_1(t)} \wedge  \overbrace{\exp(t) -1\leq 0}^{\texttt{P}_2(t)}) \nonumber
\end{align}}
where:
{\footnotesize
\begin{align}
& \exp(t)=  && \overbrace{+783.3 \cdot \omega_e\_\texttt{x}(t)}^{\texttt{T1}} 
    \overbrace{-332.6 \cdot \omega_e\_\texttt{y}(t)}^{\texttt{T2}} 
    \overbrace{+3.5 \cdot \omega_e\_\texttt{z}(t)}^{\texttt{T3}}  & \nonumber\\
 & &&     -50.57 \cdot \omega_e\_\texttt{x}(t) \cdot \omega_e\_y(t) -4751.8 \cdot \omega_e\_x(t) \cdot \omega_e\_z(t) &\nonumber\\	
& && +3588.7 \cdot \omega_e\_y(t) \cdot \omega_e\_z(t) & \nonumber \\
& && +1000 \cdot \texttt{Rwh}\_\texttt{y}(t) \cdot \omega_e\_z(t)
    -1000 \cdot \texttt{Rwh}\_\texttt{z}(t)\cdot \omega_e\_\texttt{y}(t) & \nonumber\\
 & &&     \underbrace{-54.8 \cdot \omega_e\_y(t)^2}_{\texttt{T9}}	
    \underbrace{+54.8 \cdot \omega_e\_z(t)^2}_{\texttt{T10}} & \nonumber
\end{align}}

Assumption $\texttt{A}_1$ constrains the values of the following variables within the time interval of $[0,1]$\SI{}{\second}: the estimated speed of the satellite  ($\omega_e$) and the angular momentum of the reaction wheel ($\texttt{Rwh}$) over the x-axis ($\omega_e\_\texttt{x}$ and $\texttt{Rwh}\_\texttt{x}$), the y-axis, ($\omega_e\_\texttt{y}$ and $\texttt{Rwh}\_\texttt{y}$) and the z-axis ($\omega_e\_\texttt{z}$ and $\texttt{Rwh}\_\texttt{z}$) of the body frame.
This is done by forcing the value of $\exp$ to be between $-1$ and $1$. 
Predicates $\texttt{P}_1(t)$ and  $\texttt{P}_2(t)$ are  composed of the ten terms \texttt{T1}, \texttt{T2}, \ldots, \texttt{T10} of $\exp$ and the constants $1$ and $-1$.
For example, 
the term \texttt{T3} of  $\texttt{P}_1(t)$  is 
$ +3.5 \cdot \omega_e\_\texttt{z}(t) $.

Assumption $\texttt{A}_1$ is complex, and cannot be learned by our earlier work EPIcuRus~\cite{epicurus20 }  since it is a complex function that combines three input signals of $\omega_e$ and \texttt{Rwh} with arithmetic operators.

Requirement $\phi_2$ in Table~\ref{tab:requirements} constrains the magnetic moment commanded to the magnetorquers to be in the range $[-15,15]$\SI{}{\ampere\square\meter}.
Requirement $\phi_3$ constrains the current applied to the magnetorquers to be in the range  $[-0.176,0.176]$\SI{}{\ampere}.
Finally, requirement $\phi_4$ constrains the torque acceleration applied to each of the reaction wheels to be in the range $[-0.021,0.021]$\SI{}{\meter\per\second\squared}.
Those requirements were provided by the  manufacturers of the reaction wheel and magnetorquer (see for example~\cite{acquatella2018fast}).
 According to the design documents, we expected these requirements to be satisfied for all possible input signals, i.e., without the need of adding any assumption.

\textbf{Objective.} 
Without accounting for assumption $\texttt{A}_1$,  we may falsely conclude that the \texttt{AC} model is faulty as it does not satisfy $\phi_1$. However, after restricting the inputs of \texttt{AC} with an appropriate assumption,
we can show that it satisfies $\phi_1$. Hence, there is no fault in the internal algorithm of \texttt{AC}.

In this paper, we extend EPIcuRus to provide an automated approach to infer \emph{complex} environment assumptions for system components such that they, after being constrained by the assumptions, satisfy their requirements. Our extension is applicable under the  pre-requisites \textbf{Prerequisite-1}, \textbf{Prerequisite-2}, and \textbf{Prerequisite-3} of EPIcuRus that are summarized in the following.

\textbf{Prerequisite-1.} \emph{The component \texttt{M} to be analyzed  is specified in the Simulink\textsuperscript{\tiny\textregistered} language.}
Simulink\textsuperscript{\tiny\textregistered}~\cite{Chaturvedi2009MSS1823037,Simulink} is a well-known and widely-used language for specifying the behavior
of cyber-physical systems such as those used in the automotive and aerospace domains. 
 Each Simulink\textsuperscript{\tiny\textregistered} model has a number of inputs and outputs. We denote a test input for \texttt{M} as  $\inputs=\{\inputsignal_1, \inputsignal_2 \ldots \inputsignal_m\}$ where each $\inputsignal_i$ is a signal for an input of \texttt{M}, and a test output for \texttt{M} as $\outputs=\{\outputsignal_1, \outputsignal_2 \ldots \outputsignal_n\}$ where each $\outputsignal_i$ is a signal for some output of \texttt{M}.  Simulink\textsuperscript{\tiny\textregistered} models can be executed using a simulation engine that receives  a model \texttt{M} and a test input $\inputs$ consisting of signals over a time domain $\timedomain$,
 and computes  the test output $\outputs$ consisting of signals over the same time domain $\timedomain$. 
 A \emph{time domain} $\timedomain=[0,b]$ is a non-singular bounded interval of \real .  
A \emph{signal} is a function $f: \timedomain \rightarrow \real$.
A \emph{simulation}, denoted by $\texttt{H}(\inputs,\texttt{M})=\outputs$, 
receives a test input $\inputs$ 
and produces a test output $\outputs$.

\begin{figure}[t]
\centering
\subfloat[Inputs]{
\includegraphics[width=3.5in]{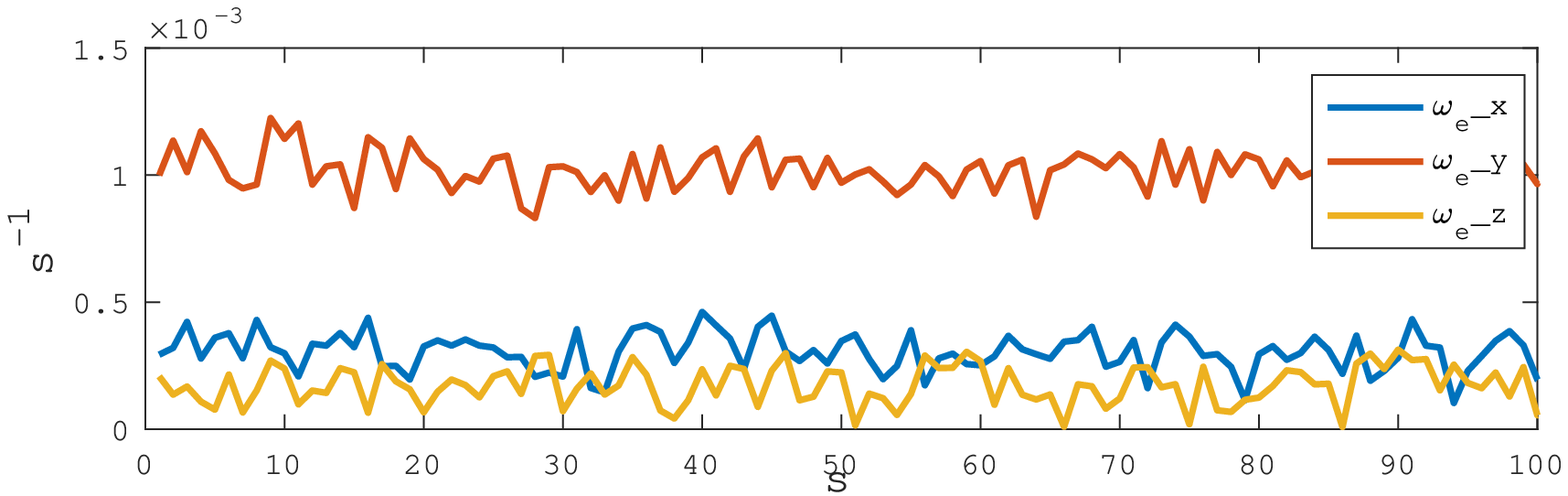}\label{fig:in}}\\
\subfloat[Outputs]{
\includegraphics[width=3.5in]{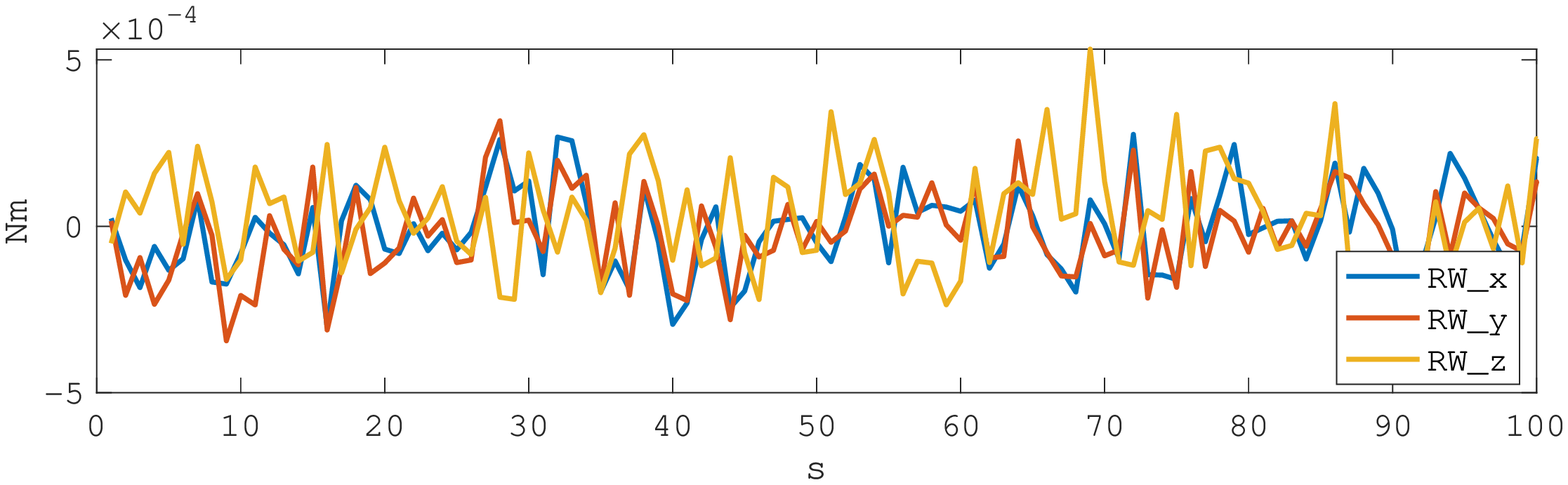}\label{fig:out}}
\caption{Example of Input/Output signals of the \texttt{AC} model.}
\label{fig:signals}
\end{figure}

For example, Fig.~\ref{fig:in} shows the three input signals of $\omega_e$, i.e., the estimated angular velocity of the satellite, over the time domain~$[0, 100]$\SI{}{\second}, for the \texttt{AC} component. Fig.~\ref{fig:out} shows three output signals obtained by simulating the model with these inputs, and representing the torque command~(\texttt{RW}) \texttt{AC} applied to the reaction wheels of the satellite over the time domain~$[0, 100]$\SI{}{\second}. 
Note that, when the behavior of a single component is analyzed, and the component is software, engineers significantly reduce the time domain.
Indeed, software components promptly react to input changes. For example, the frequency of execution of \texttt{AC} is \SI{8}{\hertz}, i.e., \texttt{AC} is executed every \SI{125}{\milli\second}. Therefore, when \texttt{AC} is evaluated, $[0, 1]$\SI{}{\second} is a reasonable time domain, since it allows executing the control logic of \texttt{AC} eight times.\footnote{The simulation time step is $\nicefrac{0.125}{4}$\SI{}{\milli\second}.}
Furthermore, the inputs can be considered  constant within this input domain.

\textbf{Prerequisite-2.} \emph{The requirement $\phi$ the component has to satisfy is specified in a logical language.} This is to ensure that the requirements under analysis can be evaluated by model checkers or converted into fitness functions required by search-based testing. Both model checking and search-based testing are part of EPIcuRus.

\textbf{Prerequisite-3.} \emph{The satisfaction of the requirements of interest over the component under analysis can be verified using a model checker.} 
In this work, we consider QVTrace~\cite{QVtrace} to exhaustively check 
whether a model \texttt{M} satisfies the requirement $\phi$ under the assumption $\texttt{A}$, i.e.,
$\langle \texttt{A} \rangle \texttt{M} \langle  \phi \rangle$. QVTrace takes as input a Simulink\textsuperscript{\tiny\textregistered} model and a requirement specified in QCT which is a logical language based on a fragment of  first-order logic. In addition, QVtrace allows users to specify assumptions using QCT, and to verify whether a given requirement is satisfied for all the possible inputs that satisfy those assumptions. QVtrace uses SMT-based model checking (specifically Z3 BMC~\cite{de2008z3}) to verify Simulink\textsuperscript{\tiny\textregistered} models. The QVTrace output can be one of the following: (1)~\emph{No violation exists} indicating that the assumption is valid (i.e., $\langle \texttt{A} \rangle \texttt{M} \langle \phi \rangle$ holds); (2)~\emph{No violation exists for $0 \leq k \leq k_{max}$.} The model satisfies the given requirement and assumption in the time interval $[0,k_{max}]$. However, there is no guarantee that a violation does not occur after $k_{max}$; (3)~\emph{Violations found} indicating that the assumption $\texttt{A}$ does not hold on the model $\texttt{M}$; and (4)~\emph{Inconclusive} indicating that QVTrace is not able to check the validity of $\texttt{A}$ due to scalability and incompatibility issues.  
\section{Assumption Generation Problem}
\label{sec:problemDef}
\IEEEPARstart{I}{n} this section, we recall the definition of the assumption generation problem for Simulink\textsuperscript{\tiny\textregistered} models introduced by Gaaloul et al.~\cite{epicurus20}.  Let \texttt{M} be a Simulink\textsuperscript{\tiny\textregistered} model. An \emph{assumption} \texttt{A} for \texttt{M} constrains the inputs of \texttt{M}. Each assumption \texttt{A} is represented as a disjunction ($\texttt{C}_1 \vee \texttt{C}_2 \vee \ldots \vee \texttt{C}_n$) of one or more constraints in $\mathcal{C}=\{\texttt{C}_1$, $\texttt{C}_2$,\ldots $,\texttt{C}_n\}$.
Each constraint in $\mathcal{C}$ is a 
first-order formula in the following 
form: 
\begin{center}
\footnotesize
$\forall \texttt{t} \in [ \tau_1, \tau_1' ]: \texttt{P}_1\texttt{(t)}  \wedge \forall t \in [ \tau_2, \tau'_2 ]: \texttt{P}_2\texttt{(t)}  \wedge \ldots \wedge \forall t \in [ \tau_n, \tau'_n ]: \texttt{P}_n\texttt{(t)}  $
 \end{center}
 
 where each $\texttt{P}_i\texttt{(t)} $ is a predicate over the model input variables, and each $[ \tau_i, \tau'_i ] \subseteq \mathbb{T}$ is a time domain. 
Recall from Section~\ref{sec:running} that  $\mathbb{T}=[0,1]$s is the time domain used to simulate the model~\texttt{M}.
 An example constraint for the \texttt{AC} model  is  the  constraint $\texttt{C}_1$ defined as follows:
\begin{center}
\footnotesize
$\texttt{C}_1::=\forall t \in 
[0,1]:(1.0 \cdot \omega_e\_\texttt{x}(t) + 1.0 \cdot \omega_e\_\texttt{y}(t) -0.0011) \leq 0$
\end{center}

$\texttt{C}_1$ constrains the sum of the values of two input signals $\omega_e\_\texttt{x}(t)$ and $\omega_e\_\texttt{y}(t)$ of the input $\omega_e$ of the \texttt{AC} model over the time domain $[0,1]$s. These signals represent, respectively, the angular speed of the satellite over the $\texttt{x}$ and $\texttt{y}$ axes of the body frame.

Let \inputs\ be a test input  for a Simulink\textsuperscript{\tiny\textregistered} model $\texttt{M}$, and let  $\texttt{C}$ be a constraint over the inputs of $\texttt{M}$. We write $\inputs \models \texttt{C}$ to indicate that the input $\inputs$ satisfies the constraint $\texttt{C}$. For example, the input $\inputs$ for the \texttt{AC} model, which is described in Fig.~\ref{fig:signals},  satisfies the constraint $\texttt{C}_1$. Note that for Simulink\textsuperscript{\tiny\textregistered} models, test inputs are described as functions over a time domain $\mathbb{T}$, and similarly, we define constraints $\texttt{C}$ as a conjunction of predicates over the same time domain or its subdomains. 

Let $\texttt{A}=\texttt{C}_1\vee \texttt{C}_2\vee \ldots \vee \texttt{C}_n$ be an assumption for model $\texttt{M}$, and let $\inputs$  be a test  input for $\texttt{M}$. The input $\inputs$ satisfies the assumption $\texttt{A}$ if $\inputs \models \texttt{A}$.  For example, consider the assumption 
\begin{align}
\footnotesize
\texttt{A}_2=\texttt{C}_1\vee \texttt{C}_2 \nonumber
\end{align}
of $\texttt{AC}$ where:
\begin{flushleft}
\footnotesize
$\texttt{C}_1::=\forall t \in [0,1]:(1.0 \cdot \omega_e\_\texttt{x}(t) + 1.0 \cdot \omega_e\_\texttt{y}(t) -0.0011) \leq 0$\\
$\texttt{C}_2::=\forall t \in [0,1]:(1.0 \cdot \omega_e\_\texttt{x}(t) + 1.0 \cdot \omega_e\_\texttt{y}(t)+ 1 \cdot \omega_e\_\texttt{z}(t) -0.0025) \leq 0$
\end{flushleft}

The input $\inputs$ in Fig.~\ref{fig:signals}  satisfies the assumption $\texttt{A}_2$ since it satisfies the constraints $\texttt{C}_1$ and $\texttt{C}_2$.

Let $\texttt{A}$  be an assumption,  and let $\mathcal{U}$ be the set of all  possible test inputs of $\texttt{M}$. We say $\texttt{U} \subseteq \mathcal{U}$ is a
 \emph{valid input set} of $\texttt{M}$ restricted by the assumption $\texttt{A}$ if for every input $\inputs \in \texttt{U}$, we have  $\inputs \models \texttt{A}$.
Let $\phi$ be a requirement for $\texttt{M}$ that we intend to verify. For every test input $\inputs$ and its corresponding test output  $\outputs$, 
we denote as $\llbracket \inputs, \outputs, \phi \rrbracket$  the degree of violation or satisfaction of $\phi$ when $\texttt{M}$ is executed for test input \inputs. 
Specifically, following existing research on search-based testing of Simulink\textsuperscript{\tiny\textregistered} models~\cite{socrates,menghi2019approximationrefinement,staliro} 
we define the degree of violation or satisfaction  as a function that  returns a value between $[-1,1]$ such that a negative value indicates that the test inputs $\inputs$ reveals a violation of $\phi$ and a positive or zero value implies that the test input  $\inputs$ is passing (i.e., does not show any violation of~$\phi$).  The function  $\llbracket \inputs, \outputs, \phi \rrbracket$   allows us to distinguish between different degrees of satisfaction and failure. When  $\llbracket \inputs, \outputs, \phi \rrbracket$   is positive, 
the higher the value, the more requirement $\phi$ is satisfied, the lower the value the more requirement $\phi$ is close to be violated. Dually, when $\llbracket \inputs, \outputs, \phi \rrbracket$   is negative, a value close to $0$ shows a less severe violation than a value  close to~$-1$.

\begin{definition}
\label{def:validinputset}
Let $\texttt{A}$  be an assumption, let $\phi$ be a requirement for $\texttt{M}$.
 Let  $\texttt{U} \subseteq \mathcal{U}$ be a 
 valid input set of $\texttt{M}$ restricted by assumption $\texttt{A}$.
We say the degree of satisfaction of the requirement $\phi$  over model $\texttt{M}$ restricted by the assumption $\texttt{A}$ is $v$, i.e., $\langle \texttt{A} \rangle \texttt{M} \langle \phi \rangle = v$, if
\begin{center}
$v=\underset{\inputs \in \texttt{U}}{min}\  \llbracket \inputs, \outputs, \phi \rrbracket  $
\end{center} 
where \outputs\ is the test output generated by the test input $\inputs \in \texttt{U}$. 
\end{definition}

\begin{definition}
\label{def:sound}
We say an assumption $\texttt{A}$ is $v$-sound\footnote{Called $v$-safe in our previous work~\cite{epicurus20}.} for a model $\texttt{M}$ and its requirement $\phi$, if $\langle \texttt{A} \rangle \texttt{M} \langle \phi \rangle>v$.
\end{definition}

 As discussed earlier,  we define the  function such that a  value $v$ larger than or equal to $0$ indicates that the requirement under analysis  is satisfied. Hence,  when an assumption $\texttt{A}$ is $0$-sound (or sound for short), 
the model $\texttt{M}$ restricted by $\texttt{A}$ satisfies $\phi$.

For a given model $\texttt{M}$,  a requirement $\phi$ and a given  value~$v$, we may have several assumptions that are $v$-sound. We are typically interested in identifying the $v$-sound assumption that leads to the largest valid input set $\texttt{U}$, and hence is less constraining.
Let $\texttt{A}_1$ and $\texttt{A}_2$ be two different $v$-sound assumptions for a model $\texttt{M}$ and its requirement $\phi$, and let $\texttt{U}_1$ and $\texttt{U}_2$ be the valid input sets of $\texttt{M}$ restricted by the assumptions $\texttt{A}_1$ and $\texttt{A}_2$.
 In our previous work~\cite{epicurus20}, we defined $\texttt{A}_1$ to be more informative than $\texttt{A}_2$ if $\texttt{A}_2 \Rightarrow \texttt{A}_1$. However, this definition only allows to compare assumptions when there exists a logical implication between the two. To enable a wider comparison among assumptions, in this work 
 we say that $\texttt{A}_1$ is more \emph{informative} than $\texttt{A}_2$ if $|\texttt{U}_1|>|\texttt{U}_2|$ where $|\texttt{U}_1|$ and $|\texttt{U}_2|$ are, respectively, the number of inputs in the sets $\texttt{U}_1$ and $\texttt{U}_2$. 
 Note that computing the size of valid test inputs for our assumptions which are first-order formulas is in general infeasible. As we will discuss in Section~\ref{sec:ev}, we provide an approximative  method to compare the size of valid test inputs to be able to compare a given pair of assumptions based on our proposed informativeness measure.  

In practical applications, there is an intrinsic tension between informativeness and soundness. 
The more informative are the assumptions,
the less effective are exhaustive verification tools in proving their soundness.
For example, QVtrace returns an inconclusive verdict for the requirement $\phi_1$ when the informative assumption $\texttt{A}_1$ (see Section~\ref{sec:running}) is considered, i.e., the tool is neither able to prove that $\phi_1$ nor to provide a counterexample showing that it does not hold. 
On the other hand, QVtrace can prove that the assumption $\texttt{A}_2$ is sound.
Therefore, in industrial applications, users should find a practical \emph{tradeoff} between informativeness and soundness. 
We will evaluate this tradeoff for our microsatellite case study in Section~\ref{sec:ev}.
 
 In this paper, provided with a model~$\texttt{M}$,  a requirement~$\phi$ and a desired  value~$v$, our goal is to generate the weakest (most informative)  $v$-sound assumption. We note that our approach, while guaranteeing the generation of $v$-sound assumptions,  does not guarantee that the generated assumptions are the most informative. Instead, we propose heuristics to maximize the chances of generating the most informative assumptions and evaluate our heuristics empirically in Section~\ref{sec:ev}.
 
\section{EPIcuRus Overview}
\label{sec:background}
\begin{figure}[t]
\includegraphics[width=\columnwidth]{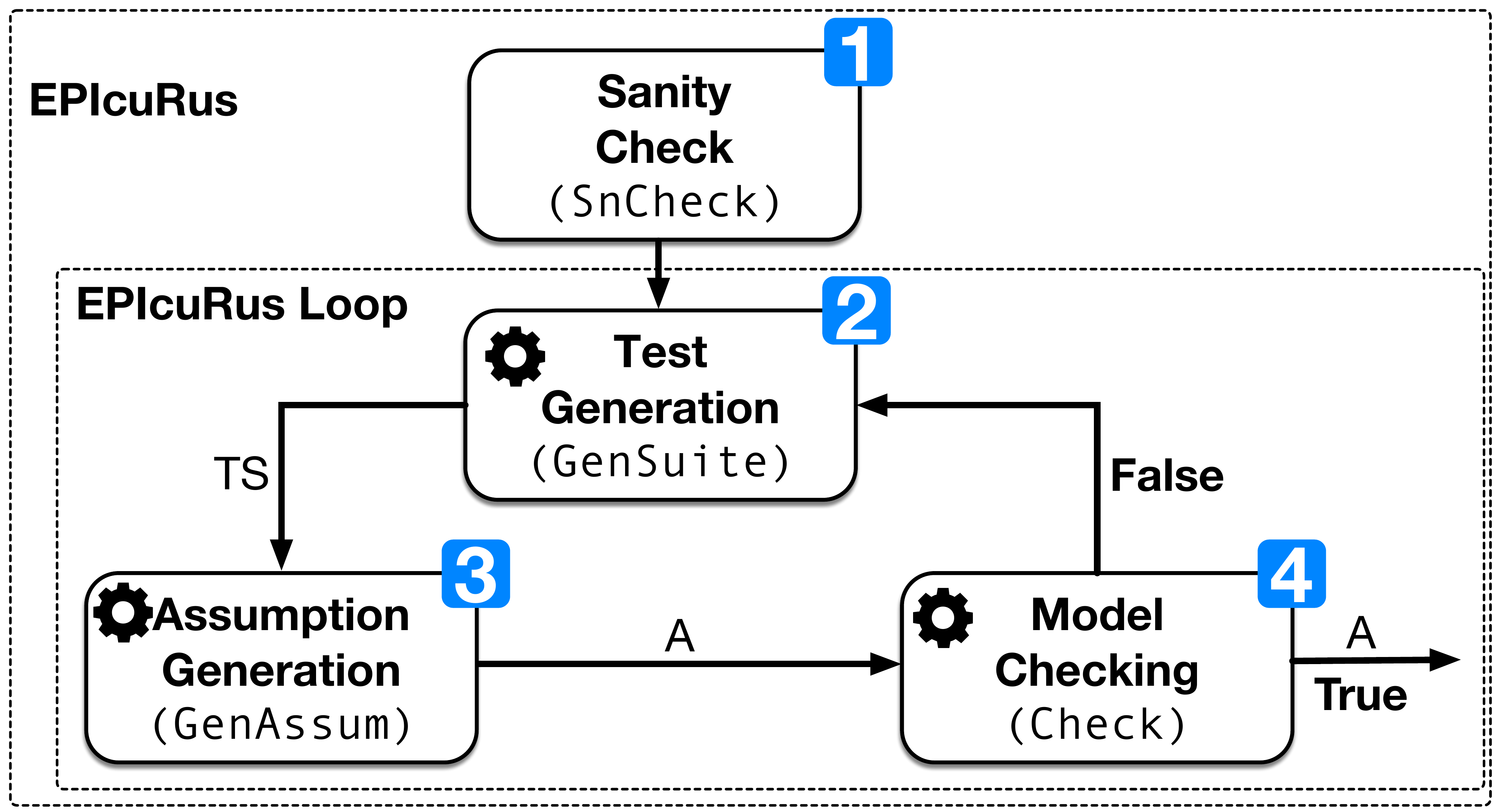}
\caption{EPIcuRus framework overview. }
\label{fig:epicurus}
\end{figure}

\IEEEPARstart{E}{PIcuRus} aims at solving the assumption generation problem. 
Fig.~\ref{fig:epicurus} shows an overview of EPIcuRus, which takes as input a Simulink\textsuperscript{\tiny\textregistered} model \texttt{M} and a requirement $\phi$, and computes an assumption ensuring that the model \texttt{M} satisfies the requirement $\phi$ when the assumption holds.
The EPIcuRus components are reported in boxes labeled with blue background numbers. 
The \emph{sanity check} (\phase{1}) verifies whether the requirement $\phi$ is satisfied (or violated) on \texttt{M} for all the inputs and therefore the assumption should not be computed. 
If the requirement is neither satisfied nor violated for all inputs, EPIcuRus iteratively performs the three steps of the EPIcuRus Loop (see 
Algorithm~\ref{alg:epicurus}) discussed  in the following:\\
\phase{2}~\emph{Test generation} (\Call{GenSuite}{}):  returns a test suite \texttt{TS} of test cases that exercise \texttt{M} with respect to  requirement $\phi$. The goal is to generate a test suite \texttt{TS} that includes both passing (i.e., satisfying $\phi$) and failing (i.e., violating $\phi$) test cases;\\
\phase{3}~\emph{Assumption generation} (\Call{GenAssum}{}):  uses the test suite \texttt{TS} to compute an assumption \texttt{A} such that \texttt{M} restricted by \texttt{A} is likely to satisfy $\phi$;\\ 
\phase{4}~\emph{Model checking} (\Call{Check}{}): checks whether \texttt{M} restricted by \texttt{A} satisfies $\phi$. We use the notation $\langle \texttt{A} \rangle \texttt{M} \langle \phi \rangle$ (borrowed from the compositional reasoning literature~\cite{DBLP:conf/tacas/CobleighGP03}) to indicate that \texttt{M}  restricted by \texttt{A} satisfies $\phi$.  If our model checker can assert $\langle \texttt{A} \rangle \texttt{M} \langle \phi \rangle$, a sound assumption is found. 

There are two stopping criteria that can be selected for  EPIcuRus.
The first  stopping criterion stops EPIcuRus whenever  the model checker can assert $\langle \texttt{A} \rangle \texttt{M} \langle \phi \rangle$. 
The second stopping criterion constrains the  timeout and allows EPIcuRus to refine the computed assumption over  consecutive iterations.
The impact of the selection of the stopping criteria on the assumption produced by EPIcuRus is discussed in Section~\ref{sec:problemDef}.

A high-level description of each of these steps is presented in the following, a detail description of the assumption generation procedures proposed in this work is provided in Section~\ref{sec:impl}.

\begin{algorithm}[t]
\caption{EPIcuRus Loop.}
\label{alg:epicurus}
\begin{flushleft}
\textbf{Inputs}. \texttt{M}: the Simulink\textsuperscript{\tiny\textregistered} model \hfill \\
 \hspace{1cm}  $\phi$: requirement of interest\\
 \hspace{1cm} \texttt{opt}: options\\
 \textbf{Outputs}. \texttt{A}: assumption
 \end{flushleft}
\begin{algorithmic}[1]
\Function{\texttt{A}=EPIcuRusLoop}{\texttt{M}, $\phi$, \texttt{opt}}
\State \hspace{-0.5cm} \texttt{TS=[]}; \texttt{A=}\texttt{null}; \Comment{Variables Initialization}
\State  \hspace{-0.5cm} \textbf{do}
\State \hspace{-0.5cm}  \hspace{0.05cm} \texttt{TS=}\Call{GenSuite}{\texttt{M}, $\phi$, \texttt{TS}, \texttt{opt}} \Comment{Test  Generation}
\State \hspace{-0.5cm} \hspace{0.05cm} \texttt{A=}\Call{GenAssum}{\texttt{TS},\texttt{opt}}; \Comment{Assum.Gen.} \label{ln:assumption}
\State \hspace{-0.5cm} \hspace{0.05cm} \texttt{A=} \Call{Check}{\texttt{A}, \texttt{M}, $\phi$} \Comment{Model Checking}
\State \hspace{-0.5cm} \textbf{while} \textbf{not} \texttt{opt.Stop$\_$Crt}
\State \hspace{-0.5cm} \Return \texttt{A};
\EndFunction
\end{algorithmic}
\end{algorithm}

\subsection{Sanity Check}
The sanity check (\phase{1}) verifies whether the requirement $\phi$ is satisfied or violated for all the inputs.
To do so, we use a model checker to respectively verify whether $\langle \top \rangle \texttt{M} \langle \phi \rangle$ or  $\langle \top \rangle \texttt{M} \langle \neg \phi \rangle$ is true.
We use the symbol  $\top$ to indicate that no assumption is considered. 
If  the requirement is satisfied for all inputs, no assumption is needed.
If the requirement is violated for all inputs, then the model is faulty, and an assumption cannot be computed as there is no input that satisfies the requirement.
The requirement \emph{passes} the sanity check if some inputs satisfy $\phi$ while others violate it, i.e., the requirement is neither satisfied nor violated for all inputs. 
In that case, the EPIcuRus loop is executed to compute an assumption.

We use QVtrace to implement our sanity check.
 QVtrace exhaustively verifies whether a Simulink\textsuperscript{\tiny\textregistered} model \texttt{M} satisfies a requirement $\phi$ expressed in the QCT language, for all the inputs that satisfy the assumption \texttt{A}, i.e., $\langle \texttt{A} \rangle \texttt{M} \langle \phi \rangle$. 

\begin{itemize}
\item  To check whether the model~\texttt{M} satisfies the requirement~$\phi$ for all inputs, we  check whether $\langle \top \rangle \texttt{M} \langle \phi \rangle$. QVTrace generates four kinds of outputs (see  Section~\ref{sec:running}).  When it returns ``No violation exists'', or ``No violation exists for 0 $\leq$ k $\leq$ k$_{max}$'', we conclude that the model under analysis satisfies the given formal requirement $\phi$ without the need to consider assumptions.
\item  To check whether the model \texttt{M} violates  requirement $\phi$  for all inputs, we check  $\langle \top \rangle \texttt{M} \langle \neg \phi \rangle$. 
 If QVtrace returns ``No violation exists'', or ``No violation exists for 0 $\leq$ k $\leq$ k$_{max}$'', we conclude that since $\neg \phi$ is satisfied, the model under analysis does not show any behavior that satisfies the requirement $ \phi$. 
 Thus, the requirement $ \phi$ is violated for any possible input, and the model is  faulty. 
\end{itemize}

 In the two previous cases, EPIcuRus provides the user with a value indicating that all the outputs of the model either satisfy or violate $\phi$.
Otherwise, the EPIcuRus loop is executed to compute an assumption.

\subsection{Test Generation}    
\label{sec:testgeneration}
 The goal of the test generation step (\phase{2}) is to generate a test suite $\texttt{TS}$ of test cases for \texttt{M} such that some test inputs lead to the violation of $\phi$ and some lead to the satisfaction of~$\phi$.  Note that, while inputs that satisfy and violate the requirement of interest can also be extracted using model checkers, due to the large amount of data needed by ML to derive accurate assumptions, we rely on simulation-based testing for data generation. Further, it is usually faster to simulate models rather than to model check them.
 For example,  performing a single simulation of \texttt{AC} and evaluating the satisfaction of  $\phi_1$ on the generated output takes $0.9$s, while model  checking \texttt{AC} against $\phi_1$ takes approximately $21.06$s. Hence, given a specific time budget, simulation-based testing leads to the generation of a larger amount of data  compared to using model checking for data generation.

  We use search-based testing techniques~\cite{survey,DBLP:conf/issta/ArcuriB11,DBLP:conf/asian/ChenLM04} for test generation and rely on simulations to run the test cases. Search-based testing allows us to guide the generation of test cases in very large search spaces. It further provides the flexibility to tune and guide the generation of  test inputs based on the needs of our learning algorithm. For example, we can use an explorative search strategy if we want to sample  test inputs uniformly or we can use an exploitative strategy if our goal is to generate more test inputs in certain areas of the search space.  For each generated test input, the underlying Simulink\textsuperscript{\tiny\textregistered} model is executed to compute the output.  
To generate input signals, we use the approach of Matlab~\cite{signals19,pareto18} that encodes signals using some parameters. 
Specifically, each signal $\inputsignal_\inputsignal$ in \inputs\ is captured by an input profile $\langle \mathit{int}_\inputsignal, R_\inputsignal, n_\inputsignal \rangle$, where $\mathit{int}_\inputsignal$ is the interpolation function,
$R_\inputsignal \subseteq \real$ is the input domain, and   $n_\inputsignal$ is the number of control points.
We assume that the number of control points is equal for all the input signals.
Provided with the values for these three parameters, we generate a signal over time domain $\mathbb{T}$ as follows: 
\begin{enumerate}
    \item we generate $n_\inputsignal$ control points, i.e., $c_{\inputsignal,1}$, $c_{\inputsignal,2}$, \ldots, $c_{\inputsignal,n_\inputsignal}$, 
equally distributed over the time domain $\mathbb{T}=[0,b]$, i.e., positioned at a fixed time distance $I=\frac{b}{n_\inputsignal-1}$.
Let $c_{x,y}$ be a control point, $x$ is the signal the control point refers to, and $y$ is the  \emph{position} of the control point. 
The control points $c_{\inputsignal,1}$, $c_{\inputsignal,2}$, \ldots, $c_{\inputsignal,n_\inputsignal}$ respectively contain the values of the signal $\inputsignal$ at time instants $0,I,2\cdot I, \ldots ,(n_\inputsignal-2) \cdot I, (n_\inputsignal-1) \cdot I$;
\item  we assign randomly generated values within the domain $R_\inputsignal$ to each control point $c_{\inputsignal,1}$, $c_{\inputsignal,2}$, \ldots, $c_{\inputsignal,n_\inputsignal}$; and  
\item we use the interpolation function $\mathit{int}_\inputsignal$ to generate a signal that connects the control points.
 The interpolation functions provided by Matlab include, among others, linear, piecewise constant and piecewise cubic interpolations, but the user can also define custom interpolation functions.
 \end{enumerate}
 To generate realistic inputs, the engineer should select an appropriate value for the number of control points ($n_\inputsignal$) and choose an interpolation function that describes with reasonable accuracy the overall shape of the input signals for the model under analysis.
Based on these inputs, the test generation procedure has to select which values $c_{\inputsignal,1}$, $c_{\inputsignal,2}$, \ldots, $c_{\inputsignal,n_\inputsignal}$ to assign to the control points for each input  $\inputsignal_\inputsignal$.

 The verdict of the requirement of interest ($\phi$) is then evaluated by 
 (i)~using the values assigned to the control points and the interpolation functions to generate input signals $\inputs$;
 (ii)~simulating the behavior of the model for the generated input signals and recording the output signals $\outputs=\texttt{H}(\inputs,\texttt{M})$; 
 (iii)~evaluating the degree $\llbracket \inputs, \outputs, \phi \rrbracket$ of satisfaction  of $\phi$ on the output signals; and
 (iv)~labelling the test case with a verdict value (\emph{pass} or \emph{fail}) depending on whether $\llbracket \inputs, \outputs, \phi \rrbracket$ is greater (or equal) or lower than zero.

The test generation step returns a test suite \texttt{TS} containing a set of test cases, each of which containing the values assigned to the control points of each input signal  and the verdict value.
Depending on the algorithm used to learn the assumption, EPIcuRus may or may not reinitialize the test suite \texttt{TS}  at each iteration. In the latter case, the test cases that were generated in previous iterations remain in the new test suite \texttt{TS} that is also expanded with new test cases.

\subsection{Assumption Generation}
Given a requirement $\phi$ and a test suite \texttt{TS}, the goal of the assumption generation step is to infer an assumption \texttt{A} such that \texttt{M} restricted based on \texttt{A} is likely to satisfy $\phi$.  
We use ML to derive an assumption based on test inputs labelled by binary verdict values.
Specifically, the assumption generation procedure infers
an assumption by learning patterns from the test suite data (\texttt{TS}). 
This is done by (i)~running the ML algorithm that  extracts an assumption defined over the control points of the input signals of the model under analysis; and (ii)~transforming the assumption defined over the values assigned to the control points into an assumption defined over the values of the input signals such that it can be checked by QVTrace. 

Different ML techniques used in the assumption generation step lead to different versions of EPIcuRus. In this work, we consider Decision Trees (DT), Genetic Programming (GP), and Random Search (RS) as alternative assumption generation policies. 
DT was recently used by Gaaloul et al.~\cite{epicurus20}, while  GP, and RS are described in Section~\ref{sec:impl} and are part of the contributions of this work.

\subsection{Model Checking} 
This step checks whether the assumption \texttt{A} generated by the assumption generation step is $v$-sound. Note that the ML technique used in the assumption generation step, being a non-exhaustive learning algorithm, cannot ensure that assumption \texttt{A} guarantees the satisfaction of $\phi$ for \texttt{M}. Hence, in this step, we use a model checker for Simulink\textsuperscript{\tiny\textregistered} models to check whether \texttt{M},  restricted by \texttt{A}, satisfies $\phi$, i.e., whether $\langle A \rangle \texttt{M} \langle \phi \rangle$ holds. 
When QVTrace returns ``No violation exists'', or ``No violation exists for 0 $\leq$ k $\leq$ k$_{max}$'', we conclude that assumption \texttt{A} ensures that \texttt{M} satisfies requirement $\phi$.

\section{Assumption Generation Procedures}
\label{sec:impl}
\IEEEPARstart{I}{n} this section, we describe our solution for learning assumptions. For a given Simulink\textsuperscript{\tiny\textregistered} model \texttt{M}, we generate assumptions over individual control point variables of the signal inputs of \texttt{M}  (Section~\ref{sec:assumptions}). We then provide a procedure to lift the generated assumptions that are defined over control points to those defined over signal variables (Section~\ref{sec:CP2snass}). 
Our algorithm to generate assumptions uses Genetic Programming~(GP) because we want  to generate complex assumptions composed of  arbitrary linear and non-linear arithmetic formulas. In addition, we introduce a baseline algorithm using Random Search~(RS) for generating assumptions.

\begin{algorithm}[t]
\caption{Genetic Programming (GP)}
\label{alg:gpalg}
\begin{flushleft}
\textbf{Inputs}. \texttt{TS}: the test suite\\
 \hspace{1.1cm} \texttt{opt}: values of the parameters of GP (Table~\ref{tab:parameters})\\
 \textbf{Outputs}. \texttt{A}:  assumption\\
 \end{flushleft}
\begin{algorithmic}[1]
\Function{\texttt{A}=GenAssum}{\texttt{TS},\texttt{opt}}
\State  \hspace{-0.5cm} \texttt{t}$=$0;
\State \hspace{-0.5cm}
\texttt{P$_0$}=\Call{Initialize}{\texttt{TS},\texttt{opt}}; \Comment{Initialize Population}
\State \hspace{-0.5cm} \texttt{P$_0$.Fit}=\Call{Evaluate}{\texttt{TS},\texttt{P$_0$}};\Comment{Assumptions Evaluation}\label{alg:init}
\State  \hspace{-0.5cm} 
\textbf{while}  \texttt{t}$<$\texttt{opt.Gen\_Size}
\textbf{do}
\State \hspace{-0.3cm}
\texttt{Off}=\Call{Breed}{\texttt{P$_t$},\texttt{opt}};\Comment{New Offspring}
\label{alg:genpop}
\State \hspace{-0.3cm} \texttt{Off.Fit}=\Call{Evaluate}{\texttt{TS},\texttt{Off}};\Comment{Evaluation}
\label{alg:eval}
\State \hspace{-0.3cm} \texttt{P$_{t+1}$}=\texttt{Off};
\Comment{New Population}
\State \hspace{-0.3cm} \texttt{t}=\texttt{t+1};
\State \hspace{-0.5cm} \textbf{endwhile}  
\State \hspace{-0.5cm} \texttt{A}=\Call{BestAssum}{\texttt{P$_0$,\ldots,P$_{t+1}$}};\Comment{Get Best Assum.}
\label{alg:save}
\State \hspace{-0.5cm} \Return \texttt{A};
\EndFunction
\end{algorithmic}
\end{algorithm}

\begin{table*}
\caption{Parameters of EPIcuRus (EP) and its Genetic Programming algorithm (GP).}
\label{tab:parameters}
\centering
\scalebox{.9}{
\begin{tabular}{c p{1.5cm} p{6cm} p{1.5cm} p{8.5cm}}
\toprule
& \textbf{Parameter} & \textbf{Description}  & \textbf{Parameter} & \textbf{Description} \\
\midrule
\multirow{3}{*}{\textbf{EP}} & \texttt{SBA} & Search-based algorithm (\texttt{GP}, \texttt{DT}, \texttt{RS}). & 
\texttt{ST} & Simulink\textsuperscript{\tiny\textregistered} Simulation time.  \\
 & \texttt{TS$\_$Size} & The number of the generated test cases per iteration.
 & \texttt{Stop$\_$Crt} & Stopping criteria:  $v$-sound assumption found (\texttt{MC}) or timeout (\texttt{Timeout}). \\
& \texttt{Timeout} & EPIcuRus timeout. &
\texttt{Nbr$\_$Runs} & Number of experiments to be executed.\\
 \midrule
\multirow{6}{*}{\textbf{GP}}  & 
\texttt{Max$\_$Conj} & 
Maximum number of conjunctions in an assumption.
& 
\texttt{Max$\_$Disj} & Maximum number of disjunctions in an assumption.\\
&
\texttt{Const$\_$Min} & Minimum constant value.
& \texttt{Const$\_$Max} & Maximum constant value.
\\
& \texttt{Max$\_$Depth} & Maximum depth of the syntax tree. 
& \texttt{Init$\_$Ratio} &  Percentage of the assumptions copied from the last population.
\\
& \texttt{Pop$\_$Size} & Number of individuals per population. 
& \texttt{Gen$\_$Size} & Number of generations.\\
& \texttt{Sel$\_$Crt} & The selection criterion.
&
\texttt{T$\_$Size} & 
The number of individuals chosen for the tournament selection. \\
& \texttt{Mut$\_$Rate} & Probability of applying the mutation operator.
&
\texttt{Cross$\_$Rate} & Probability of applying the crossover operator.\\
\bottomrule
\end{tabular}}
\end{table*}

\subsection{Learning  Assumptions on Control Points with GP}
\label{sec:assumptions}
Genetic programming (GP) is a technique for evolving programs from an initial randomly generated population in order to  find fitter programs (i.e., those optimizing a desired  fitness function).
In our work, we use Strongly Typed Genetic Programming (STGP)~\cite{koza1992genetic, montana1995strongly}, a variation of GP designed to ensure that all the individuals within a population follow a set of syntactic rules specified by a grammar. 
The steps of our GP procedure are summarized by
Algorithm~\ref{alg:gpalg}.  
First (\Call{Initialize}{}), the algorithm creates  an initial population containing a set of possible solutions (a.k.a. individuals).
A fitness measure is used to assess how well each individual solves the problem (\Call{Evaluate}{}).
In the evolutionary part,
the algorithm iteratively generates new populations (\Call{Breed}{}). It  extracts a set of parents individuals and generates  an offspring set by applying genetic operators to the parents. The algorithm then evaluates
the individuals of the offspring (\Call{Evaluate}{}) and uses the offspring set as the  new population \texttt{P}$_{t+1}$  for the next iteration.
The breeding and evaluation steps are repeated for a given number of generations (\texttt{opt.Gen\_Size}).
Then, the algorithm finds among all the individuals of the generated populations the individual with the highest fitness (\Call{BestAssum}{}).
The algorithm returns 
the individual with the highest fitness (\texttt{A}).

In the following, we describe how we use Algorithm~\ref{alg:gpalg} to generate assumptions over individual control point variables of the input signals of \texttt{M}. 

\textbf{Representation of the Individuals.} Each individual represents an assumption over individual control point variables of the input signals  of \texttt{M}. 
Specifically, assumptions are defined according to the syntactic rules of the grammar provided in Fig.~\ref{fig:assoncp}.
Furthermore, 
we constraint each arithmetic expression to contain only signal control points in the same position.
For example, the assumption
\begin{center}
    $(c_{\inputsignal_1,1} - c_{\inputsignal_2,1} - 20\le 0) \lor ((c_{\inputsignal_1,2} < 0) 
    \wedge (c_{\inputsignal_2,2}-2.5 =0))$
\end{center} is defined according to the grammar in Fig.~\ref{fig:assoncp}. It constrains the values of the control points
$c_{\inputsignal_1,1}$,  $c_{\inputsignal_2,1}$,
$c_{\inputsignal_1,2}$, and $c_{\inputsignal_2,2}$, and each  arithmetic expression contains control points that refer to the same position. For example,  $c_{\inputsignal_1,1}+c_{\inputsignal_2,1}$ contains the signal control points
$c_{\inputsignal_1,1}$,  $c_{\inputsignal_2,1}$  of the input signals
$\inputsignal_1$ and $\inputsignal_2$ in position $1$.

\begin{figure}[]
    \centering
    \begin{footnotesize}
\begin{tabular}{l@{\ }l@{\ } l@{\ } p{0.05mm} l@{\ } l@{\ } p{63mm}}
        \synt{or-exp} &::= & \synt{or-exp} $\vee$ \synt{or-exp} $|$ \synt{and-exp}  \\ 
         \synt{and-exp} &:: = &
        \synt{and-exp} $\wedge$ \synt{and-exp} $|$ \synt{rel}
        \\
        \synt{rel} &:: = &
        \synt{exp} ($<$ $|$ $\leq$ $|$ $>$ $|$ $\geq$ $|$ $=$ ) $0$\\
        \synt{exp} &:: = &
        \synt{exp} ($+|-|*|/$ ) \synt{exp} $|$ \synt{const} $|$ \synt{cp}\\
    \end{tabular}
    \end{footnotesize}
    \caption{Syntactic rules of the grammar that defines the assumptions on control points. The symbol $|$ separates alternatives, \synt{const} is a constant value, and \synt{cp} is a variable that refers to a control point.}
    \label{fig:assoncp}
\end{figure}

\textbf{Initial Population.}
The initial population contains 
a set of \texttt{Pop$\_$Size} individuals. 
The population size remains the same throughout the search.
The method \Call{Initialize}{} generates the initial population. Recall from Section~\ref{sec:background} (see Algorithm~\ref{alg:epicurus}) that \Call{GenAssum}{} is called within EPIcuRus iteratively.

The first time that \Call{Initialize}{} is called it generates an initial set of randomly generated individuals. 
We use the grow method ~\cite{poli2008field} to randomly generate each individual in the initial population. 
The grow method generates a tree with a maximum depth \texttt{Max$\_$Depth}.
It first creates the root node of the tree labeled with the Boolean operator $\lor$ or $\wedge$ or one among the relational operators $<$,$\leq$,$>$,$\geq$ and $=$.
Then, it iteratively generates the child nodes as follows.
If the node is not a terminal, the algorithm considers the production rule of the grammar in Fig.~\ref{fig:assoncp} associated with the node type. 
One among the alternatives specified on the right side of the production rule is randomly selected and used to generate the child nodes.
Then, the child nodes are considered.
If the node is a terminal,  
depending on whether the node type is \synt{const} or \synt{cp}, 
a random constant value
within the range 
$[\texttt{Const$\_$Min},\texttt{Const$\_$Max}]$
or signal control point is chosen with equal probability among the set of all the control points, respectively.
To ensure that the generated tree does not exceed the maximum depth (\texttt{Max$\_$Depth})
the algorithm constrains the type of the nodes that 
can be considered 
as the size of the tree increases. 
The algorithm also forces each arithmetic expression (\synt{exp}) to only use control points in the same position. 
When all the nodes are considered the individual is returned.

In the subsequent iterations, however, \Call{Initialize}{} copies a subset 
of individuals from the last population generated by the previous execution of \Call{GenAssum}{}.
The number of copied individuals is determined by the initial ratio (\texttt{Init$\_$Ratio}) in its initial population and then randomly generates the remaining elements required to reach the size \texttt{Pop$\_$Size}. This allows GP to reuse some of the individuals generated previously instead of starting from a fully random population each time it is executed.

\textbf{Fitness Measure}. 
Our fitness measure is used by the \Call{Evaluate}{} method to assess how  $v$-sound and informative the assumption individuals in the current populations are (see Section~\ref{sec:problemDef}). The fitness measure relies on the test cases contained in the test suite \texttt{TS}. We say a test case \texttt{tc} in \texttt{TS} satisfies an  assumption \texttt{A} if \texttt{tc} is a satisfying value assignment for \texttt{A}. 
We compute the number of passing test cases in \texttt{TS} that satisfy \texttt{A} and denote it by \texttt{TP}. 
We also compute the number of failed test cases in \texttt{TS} that satisfy \texttt{A} and denote it by \texttt{TN}. 
We then compute the $v$-sound degree of an assumption \texttt{A} as follows:
\begin{center}
$\text{\emph{v-sound}} =\dfrac{\text{\texttt{TP}}}{\texttt{TP}+\texttt{TN}}$
\end{center}
The variable \emph{v-sound} assumes a value between $0$ and $1$.  The higher the value, the more passing test cases in \texttt{TS} satisfy the assumption. 
When \emph{v-sound} = $1$, all test cases in \texttt{TS} that satisfy the assumption lead to a \emph{pass} verdict. Since there is no failing test case in \texttt{TS} that satisfies the assumption, the assumption is likely to be \emph{v-sound}.

To measure whether an assumption is informative,
we compute the ratio of test cases in \texttt{TS} satisfying the assumption (\texttt{TP}+\texttt{TN}) over the total number of test cases in TS:
\begin{center}
\text{\emph{informative}} =$\dfrac{\text{\texttt{TP}}+\text{\texttt{TN}}}{\text{\texttt{TS}}}$
\end{center}
The higher \emph{informative}, the weaker the assumption, and the more useful it is when informing engineers about when a requirement is satisfied. 

Having computed \emph{v-sound} and \emph{informative} for an assumption~\texttt{A}, we compute the fitness function for \texttt{A} as follows: 
\begin{align}
\Call{Fn}{}=\text{\emph{v-sound}} + \lfloor \text{\emph{v-sound}} \rfloor*\text{\emph{informative}} \nonumber
\end{align}
where $\lfloor \rfloor$ is the Matlab floor operator~\cite{floor}. 
    This operator returns $0$ for all the values of \emph{v-sound} within the interval $[0,1)$ and returns~$1$ if \emph{v-sound}  is equal to~$1$. 
Function \Call{Fn}{} 
returns the \emph{v-sound} value when \emph{v-sound} is within the interval $[0,1)$. 
When \emph{v-sound} is equal~$1$ it returns the value $1+ \emph{informative}$.
Intuitively,  \Call{Fn}{} starts considering the informative value only when an assumption is $v$-sound. 
This fitness function guides the search toward the detection of $v$-sound assumptions (which is our primary goal) that are as informative as possible.

We note that our fitness provides one way to combine \emph{v-sound} and \emph{informative} values that fitted our needs. Developers may identify other ways to combine these two values to prioritize either \emph{v-sound} or \emph{informative} assumptions.

\textbf{Parents Selection}. It
uses the fitness values to select parent individuals
for crossover and mutation operations (see \Call{SelectParents}{}) that will be used to generate a new population. We implemented the following standard selection criteria of GP: \emph{Roulette Wheel Selection (RWS)}, \emph{TouRnament Selection (TRS)} and \emph{Rank Selection (RS)}~\cite{luke2013essentials}.

\begin{figure}[t]
\centering
\includegraphics[width=2in]{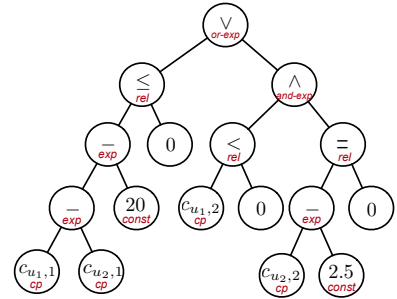}
\caption{Syntax tree associated with an individual. }
\label{fig:treeexample}
\end{figure}

\textbf{Genetic Operators.}
The genetic  operators of GP act on the syntax tree of individuals.
For example, the syntax tree of 
\begin{center}
    $(c_{\inputsignal_1,1} - c_{\inputsignal_2,1}-20 \le 0) \lor ((c_{\inputsignal_1,2} < 0) 
\wedge (c_{\inputsignal_2,2}-2.5 =0))$
\end{center} is shown in Fig.~\ref{fig:treeexample}.
Each node of the syntax tree represents a portion of the individual and is labeled (italic red label)  with the identifier of the corresponding syntactic rule of the grammar of Fig.~\ref{fig:assoncp}. 

The \Call{Breed}{} method generates an offspring by 
\begin{itemize}
    \item 
     either  applying the \emph{crossover}  operator to generate two new individuals (with probability \texttt{Cross$\_$Rate}) or randomly selecting an individual from \texttt{P$_t$}; and
    \item applying the \emph{mutation} operator (with probability \texttt{Mut$\_$Rate}) to the individuals returned by the previous step.
\end{itemize}
The \emph{crossover} and \emph{mutation} genetic operators are summarized in the following.

We use one-point crossover~\cite{onepointcrossover} as \emph{crossover} operator.
One-point crossover
(i)~randomly selects two parent individuals;
(ii)~randomly selects one subtree in each parent; and
(iii)~swaps the selected subtrees resulting in two child individuals. To ensure that the child individuals are compliant with our representation, we force the following constraints to hold:
\begin{itemize}
\item The type of the root nodes of the subtrees is the same;
\item The depth of the child individuals does not exceed \texttt{Max$\_$Depth};
\item The number of conjunctions and disjunctions of the child individuals does not exceed \texttt{Max$\_$Conj} and \texttt{Max$\_$Disj}, respectively;
\item When the type of the root nodes of the subtrees is \synt{exp}, 
all the signal control points of the subtrees are in the same positions.
\end{itemize}

We use point mutation~\cite{poli1998schema}
as a \emph{mutation} operator.
Point mutation mutates a child individual
by randomly selecting one subtree and replacing it with a randomly generated tree. 
To create the randomly generated tree
we adopt the procedure used within the \Call{Initialize}{} method.
Additionally, to ensure that the mutated child individual is compliant with our representation, our implementation ensures the constraints specified for the crossover operator are also satisfied here.

The full set of GP parameters is summarized in Table~\ref{tab:parameters}.

\textbf{Random Search.} It proceeds following the steps of Algorithm~\ref{alg:gpalg}.
However, at each iteration, a new set of individuals is randomly generated by adopting the same procedure used within the \Call{Initialize}{} method.

\subsection{Control Points-Based to Signal-Based Assumptions}
\label{sec:CP2snass}
To use assumptions in QVtrace, it is necessary to translate assumptions that constrain control point values to assumptions that constrain signal values. To do so, we proceed as follows.
 Recall that 
control points 
$c_{\inputsignal,1}$, $c_{\inputsignal,2}$, \ldots $c_{\inputsignal,n_u-1}$, $c_{\inputsignal,n_u}$ are respectively  positioned at time instants $0,I,2\cdot I, \ldots ,(n_\inputsignal-2) \cdot I, (n_\inputsignal-1) \cdot I$ and that
any arithmetic expression contains only signal control points in the same position. Each expression \texttt{exp} that constrains the values of control points in position $j$ is translated into an expression $\forall t \in [(j-1)\cdot I,j\cdot I): \texttt{exp}'$ where  $\texttt{exp}'$ is obtained by substituting each control point  $c_{\inputsignal_x,j}$ with the expression  $\inputsignal_x(t)$ modeling the input signal $\inputsignal_x$ at time~$t$.
Intuitively, this substitution specifies that the expression $\texttt{exp}$ holds within the entire time interval  $[(j-1)\cdot I,j\cdot I)$.
For example, assuming that the control points $1$, $2$ and $3$ are respectively positioned at time instant $0$, $5$ and $10$, the assumption on the control points
\begin{center}
    $(c_{\inputsignal_1,1} - c_{\inputsignal_2,1}-20 \le 0) \lor ((c_{\inputsignal_1,2} < 0) 
    \wedge (c_{\inputsignal_2,2}-2.5 =0))$
\end{center} 
is translated into an assumption over the signal variables as follows:
\begin{center}
    $
    (\forall t \in [0,5) : \inputsignal_1 (t) - \inputsignal_2(t)-20 \le 0)$\hspace{0.1cm} $\lor $\\
$((\forall t \in [0,5) :
    \inputsignal_1 (t) < 0)$\hspace{0.1cm} $ 
    \wedge $\hspace{0.1cm} $
(\forall t \in [5,10) :
\inputsignal_2 (t) - 2.5=0))$
\end{center}  
\section{Evaluation}
\label{sec:ev}
In this section, we  evaluate our contributions by answering the following research questions:\\
$\bullet$  \emph{RQ1 (Comparison of the search-based techniques). How does GP compare with DT and RS in generating informative, sound assumptions over signal variables? (Section~\ref{sec:comparison})}\\ 
To answer this question, we compared the different search-based techniques of EPIcuRus and empirically assessed
whether  GP learns sound assumptions that are more informative than the ones learned by DT and RS.
We are not aware of any tool other than EPIcuRus for computing signal-based assumptions that we could use as a baseline of comparison. 
To answer this question we relied on a public-domain set of representative models of CPS components~\cite{9218211} from Lockheed Martin~\cite{lockheedmartin}---a company working in the aerospace, defense, and security domains--- and the model of our satellite case study (\texttt{AC}).
Recall that EPIcuRus targets individual components, that can be analyzed using a model checker, and is generally not applicable to the entire industrial CPS models, such as the ADCS model (see sections~\ref{sec:intro} and~\ref{sec:running}).\\ 
$\bullet$  \emph{RQ2 (Usefulness). How useful are the assumptions learned by EPIcuRus?}\\
To answer this question, we empirically assessed whether EPIcuRus can learn assumptions that are meaningful and understandable by engineers by comparing them with assumptions engineers would normally write.
We answer this question by using 
our best search technique, according to RQ1 results, and the \texttt{AC} component of ADCS since 
\begin{enumerate*}
    \item this is a representative example of an industrial CPS component (see Section~\ref{sec:running}), and 
    \item we could interact with the engineers that developed the \texttt{AC} component to evaluate how the assumptions computed by EPIcuRus compare with the assumptions they manually wrote.
\end{enumerate*}
To answer our research question, we considered the requirement~$\phi_1$ of \texttt{AC} (see Section~\ref{sec:running}) since EPIcuRus confirmed that the requirements $\phi_2$,  $\phi_3$, and $\phi_4$, are satisfied for all  possible input signals.
Since there is, when dealing with complex components, a  \emph{tradeoff} among the informativeness of the assumptions returned by EPIcuRus and the capability of QVtrace to confirm their soundness (see Section~\ref{sec:problemDef}), engineers often have the choice to either learn highly informative assumptions, whose soundness cannot be confirmed by a solver like QVtrace, or alternatively learn simpler assumptions, which are less informative  but whose soundness can be verified exhaustively.
Our goal is to investigate such tradeoff when analyzing industrial CPS components.
Therefore, to answer RQ2, we are considering two sub-questions:
\begin{itemize}
    \item \emph{RQ2-1.  How useful are EPIcuRUS assumptions when demonstrating their soundness is not a priority? (Section~\ref{sec:usefulnessv1})}
    \item \emph{RQ2-2. How useful are EPIcuRUS assumptions when learning assumptions whose soundness can be verified is prioritized? (Section~\ref{sec:usefulnessv2})}
\end{itemize}

\emph{Implementation and Data Availability.} We extended the original Matlab implementation of EPIcuRus~\cite{epicurus20}. 
We decided to implement the procedure presented in Section~\ref{sec:impl} by reusing existing tools. Among the many tools available in the literature (e.g., Weka~\cite{frank2016weka}, GPLAB~\cite{silva2003gplab}, GPTIPS~\cite{searson2015gptips}, Matlab GP toolbox~\cite{GPMatlab,madar2005genetic}), we decided to rely on tools developed in Matlab. This facilitates the integration of our extension within EPIcuRus, and restricted our choice to GPLAB, GPTIPS, and the Matlab GP toolbox.
Among these, we implemented our techniques on the top of GPLAB. 
We chose GPLAB since it allows the introduction of new genetic operators by adding new functions. We exploited this feature to implement the genetic operators of the  procedure presented in Section~\ref{sec:impl}. Our implementation and results are publicly available~\cite{githubmaterial}.

\subsection{RQ1 ---  Comparison of the  Search-Based Techniques}
\label{sec:comparison}

\begin{table*}[t]
\caption{Identifier (\texttt{ID}), 
name, 
description, 
number of blocks (\#\texttt{Bk}), 
inputs (\#\texttt{In}), 
outputs (\#\texttt{Out}),
simulation time (\texttt{ST}),
number of requirements (\#\texttt{Reqs}),
and the number of requirements we used to answer research question RQ1, i.e., they pass the sanity check (\#\texttt{Reqs} within round brackets)
of each Simulink\textsuperscript{\tiny\textregistered} model of the components  of our study subjects.} 
\label{table:cpsmodels}
\begin{center}
\begin{tabular}{l p{2.4cm} p{8.5cm}  l l l l l   }
 \toprule   
\bf \texttt{ID} & \bf Name  & \bf Description & {\bf \#\texttt{Bk}} & {\bf \#\texttt{In}} & {\bf \#\texttt{Out}} & {\bf \texttt{ST}(s) } & {\bf \#\texttt{Reqs} }  \\
 \midrule  
\texttt{TU} & Tustin& A numeric model that computes integral over time.& 57 & 5 & 10 & 10 & 5 (2)   \\
\texttt{EB} & Effector Blender & A controller that  computes the optimal effector configuration for a vehicle. & 95 & 1 & 7 & 0 & 3 (0)\\
\texttt{SW} &  Integrity Monitor & Monitors the airspeed and checks for hazardous situations. & 164 & 7 &5 & 10 & 2 (0)\\
\texttt{FSM} & 	  	Finite State Machine   &  Controls  the autopilot mode in case of some environment hazard. & 303 & 4 & 1 & 10 & 13 (2)  \\
\texttt{REG} &	Regulator & A typical PID controller.& 308 & 12 & 5 & 10 & 10 (6) \\
\texttt{NLG} & Nonlinear Guidance &  A  guidance algorithm for an Unmanned Aerial Vehicles (UAV).& 373 & 5 & 5 & 10 & 2 (0)  \\
\texttt{TX} & Triplex  & A redundancy management system. & 481 & 5 & 4 & 10 & 4 (0) \\
 \texttt{TT} & 	Two Tanks$^\ast$ & A controller regulating the   incoming and outgoing flows of two tanks.  
  	 & 498 & 2& 11 & 14 & 32 (8)   \\
\texttt{NN} & Neural Network & A  predictor neural network model with two hidden layers. & 704 & 2 & 1 & 100 & 2 (0)  \\ 
\texttt{EU} & Euler & Computes the rotation matrices for an inertial frame in a Euclidean space.& 834 & 4 & 2 & 10 & 8 (0)\\
\texttt{AP} & Autopilot & A DeHavilland Beaver Airframe with Autopilot system. & 1549 & 7 & 1 & 1000 & 11 (0) \\
\texttt{AC} & Attitude Control & Attitude control component of the ADCS of the ESAIL micro-satellite. & 438 & 10 & 4 & 1 & 2 (1) \\
  	\bottomrule
\end{tabular}
\end{center}
$^\ast$ does not support multiple control points.
\end{table*}

To compare GP, DT, and RS, 
we considered 12  models of CPS components and 94 requirements~\cite{9218211}.
These models include 11 models developed by Lockheed Martin~\cite{lockheedmartin} and the model of our satellite case study (\texttt{AC}).
The models and requirements from Lockheed Martin were also recently used to compare model testing and model checking~\cite{NejatiGMBFW19} and to evaluate our previous version of EPIcuRus~\cite{epicurus20}. Table~\ref{table:cpsmodels} contains the description, number of blocks, inputs, and outputs of each CPS component model. 
It also contains the simulation time and the number of requirements considered for each model. 

Out of the 94 requirements, 27 could not be handled by QVtrace (violating \textbf{Prerequisite-3}).
 For 16 requirements, the Simulink\textsuperscript{\tiny\textregistered} models of the 
 CPS components were not supported by QVtrace. For 11 requirements, QVtrace returned an inconclusive verdict due to scalability issues.
For 48 of the 67  requirements that can be handled by QVtrace, EPIcuRus did not pass the sanity check:  QVtrace could prove 47 requirements and refuted one requirement. 
Therefore, to answer research question RQ1, we considered the 19 requirements, that can be handled by QVtrace,  pass the sanity check,  and required the assumption generation procedure to be executed (column  \#\texttt{Reqs} of Table~\ref{table:cpsmodels} within round brackets).

\textbf{Methodology and Experimental Setup.} To answer our research question, we  configured the parameters of GP in Table~\ref{tab:parameters} according to values in   
Table~\ref{tab:expGPconfig}. 
We chose default values  from the literature~\cite{poli2008field} for the  population size (\texttt{Pop$\_$size}), mutation rate (\texttt{Mut$\_$rate}), crossover rate (\texttt{Cross$\_$Rate}), and the max tree depth (\texttt{Max$\_$Depth}) parameters.
We set tournament selection (TRS) as selection criterion (\texttt{Sel$\_$Crt})  since, when compared with other selection techniques, it leads to populations with higher fitness values~\cite{poli2008field}. We set the value of the tournament size (\texttt{T$\_$Size}) according to the results of an  empirical study on ML parameter tuning \cite{arcuri2013parameter}. 
We set the maximum number of generations (\texttt{Gen$\_$Size}),
the number of conjuntions (\texttt{Max$\_$Conj}) and disjunctions (\texttt{Max$\_$Disj}), and the initial ratio (\texttt{Init$\_$Ratio}) based on the results of a preliminary analysis we conducted, over the considered study subjects, where we determined the average number of generations needed to reach a plateau.
We assigned to \texttt{Const$\_$Min} and \texttt{Const$\_$Max}, respectively, the lowest and highest values the input signals can assume in our study subjects. 
We set the number of tests in the test suite (\texttt{TS$\_$Size}) to $300$, which was  the value used to evaluate falsification-based testing tools in the ARCH-COMP 2019 and 2020 competitions~\cite{ernst2019arch,ARCH20}. 

\begin{table}
\begin{center}
\caption{Values for the parameters of Table~\ref{tab:parameters}  used for RQ1.}
\label{tab:expGPconfig}
\begin{tabular}{c c  c c  c }
\toprule
& \textbf{Parameter} & \textbf{Value} & \textbf{Parameter} & \textbf{Value} \\
\midrule
\multirow{3}{*}{\textbf{EP}}
& \texttt{SBA} & \texttt{DT}/\texttt{GP}/\texttt{RS} &
\texttt{ST} & see Table~\ref{table:cpsmodels}\\
& \texttt{TS$\_$Size}  & \cellcolor{orange!30} $300$ &
\texttt{Stop$\_$Crt} & \texttt{Timeout} \\
& \texttt{Timeout} & \cellcolor{purple!30} $1$h
& \texttt{Nbr$\_$Runs} & \cellcolor{orange!30} $100$\\
\midrule
\multirow{6}{*}{\textbf{GP}}
&
\texttt{Max$\_$Conj} & \cellcolor{yellow!30} $3$ or $4$
& \texttt{Max$\_$Disj} & \cellcolor{yellow!30} $2$
\\
& \texttt{Const$\_$Min} & \cellcolor{purple!30} $-100$ &
\texttt{Const$\_$Max} & \cellcolor{purple!30} $100$ \\
&
\texttt{Max$\_$Depth}& \cellcolor{blue!30} $5$ &
\texttt{Init$\_$Ratio}& \cellcolor{yellow!30} $50\%$ \\ 
& \texttt{Pop$\_$Size} & \cellcolor{blue!30} $500$ &
\texttt{Gen$\_$Size}&  \cellcolor{yellow!30} $100$ \\
&\texttt{Sel$\_$Crt} & \cellcolor{blue!30}  \emph{TRS} &
\texttt{T$\_$Size} & \cellcolor{green!30} 7\\
& \texttt{Mut$\_$Rate} & \cellcolor{blue!30} $0.1$  &
 \texttt{Cross$\_$Rate} & \cellcolor{blue!30} $0.9$ \\
\bottomrule
\end{tabular}
\end{center}
$^\ast$ The values within the  framed  boxes
\colorbox{blue!30}{ \makebox(2,2){} },
\colorbox{green!30}{
\makebox(2,2){}
}, and
\colorbox{orange!30}{\makebox(2,2){ }
} are respectively from~\cite{poli2008field},~\cite{arcuri2013parameter}, and \cite{ernst2019arch}.
The value within the framed box \colorbox{yellow!30}{
\makebox(2,2){}} is based on  a preliminary analysis of the considered study subjects. 
The values within the framed boxes \colorbox{purple!30}{\makebox(2,2){}} are selected based on  our domain knowledge on the considered study subjects.  
\end{table}

We configured RS by noticing that RS reuses part of the algorithm of GP. 
Therefore, for the parameters of RS, which are a subset of the parameters of GP, we assigned the same values considered for GP.
Finally, we configured DT by considering the same values used by our earlier work, Gaaloul et al.~\cite{epicurus20}, to evaluate EPIcuRus.

To answer RQ1, we performed the following experiment. 
We  considered each of the 19 requirements under analysis and three different input profiles  with respectively one~(\texttt{IP}), two~(\texttt{IP$^\prime$}), and three~(\texttt{IP$^{\prime\prime}$}) control points. 
We chose the number of control points of the input profiles based on the default input signals provided by the models of the CPS components. For the eight requirements  of Two Tanks~(\texttt{TT}), only the input profile \texttt{IP} was considered since this model only supports constant input signals. 
Therefore, in total, we considered $32$  requirement-profile combinations.
For each combination, we ran  EPIcuRus with GP, DT, and RS. 
We set a timeout of one hour, which is reasonable for this type of applications. 
 As  typically done in similar works (e.g.,~\cite{ernst2019arch,menghi2019approximationrefinement}),
we repeated each run $100$ times to account for  the randomness of the test case generation procedure. Therefore, in total, we executed $9600$ runs\footnote{
We executed our experiments on the HPC facilities of the University of Luxembourg~\cite{VBCG_HPCS14}.}:  
$3200$ runs ($32\times100$) for each of GP,  DT, and RS.  For each run, we recorded whether a sound assumption was returned. Furthermore, we computed the informative value associated with the assumption.
To compute an \emph{informative value} (\texttt{INF}$\_$\texttt{V}), we need to compute  the size of the valid input set for each assumption (see $|U|$ in Definition~\ref{def:validinputset}). We do so empirically. Specifically, we generate 
100 different value assignments for  control points that are uniformly distributed within the value ranges of control points. We compute an informative value for each assumption by counting the number of value assignments that satisfy the assumption. The higher this number, the more informative the assumption.

\begin{figure}[t]
\centering
\includegraphics[width=0.85\columnwidth]{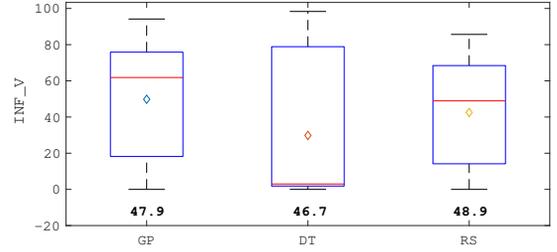}
\caption{
Comparing GP, DT, and RS. The box plots show the informative value of GP, DT, and RS (labels on the bottom of the figure). 
Diamonds depict the average.
The value below the box plot is 
 the percentage of runs, across all the runs, in which the technique was able to compute a sound assumption.}
\label{fig:rq1}
\end{figure}

\textbf{Results.} 
The results of our comparison are reported in Figure~\ref{fig:rq1}.
 Each box plot 
reports the informative value (\texttt{INF}$\_$\texttt{V}) of the assumptions computed by GP, DT, and RS, and is labelled with the percentage of runs, across the $3200$ runs, in which the technique was able to compute a sound assumption  (the value reported below the box plot).
 The average informative values of the assumptions computed by GP, DT, and RS across
 their different runs are, respectively and  approximately,  $50$, $30$ and $42$. 
  Though there are  variations across the different combinations of requirements and input profiles, GP can compute assumptions with an informative value, that is, on average, $20\%$ and $8\%$ higher than that of the assumptions computed by DT and RS, respectively. 
 
 Across all the runs, GP was able, on average, to compute a sound assumption in $47.9\%$ of the cases ($1533$ out of $3200$), while DT, and RS were able to compute a sound assumption in respectively $46.7\%$ ($1495$ out of $3200$) and $48.9\%$ ($1565$ out of $3200$) of the cases. 
Therefore, GP is, on average,  slightly more effective ($1.2\%$) than  DT, and slightly less effective than RS ($1\%$) in computing sound assumptions.
For each requirement-profile combination,
Table~\ref{tab:resultsForModel} reports the number of runs, among the $100$ runs executed for the requirement-profile combination, in which GP, DT, and RS were able to compute a sound assumption. 
When GP was less effective than DT and RS, in many runs, across all the different iterations, GP was able to learn informative assumptions that were close to actual sound assumptions, but not sound. Intuitively, in these cases, maximizing the informativeness of the assumption, by searching for assumptions with a higher fitness, leads GP away from the generation of assumptions that are sound.
For example, for the requirement ($\phi_{1}$) of Two Tanks (\texttt{TT}), GP was returning, in one of its runs, the assumption $\texttt{t1h}\le 0.5855 \lor \texttt{t1h} \geq 2.0065$ for one of the inputs of Two Tanks. 
This assumption is not sound. However, the assumption $\texttt{t1h}< 0.58 \lor \texttt{t1h} \geq 2$ is sound. 
For the same requirement, RS and DT were respectively returning,  in one of their runs, the assumptions $\texttt{t1h} \geq 2.0259$ and $\texttt{t1h}<0.5655 \lor \texttt{t1h} \geq 2.0265$  which are sound but much less informative. 
Therefore, GP learned assumptions that are the closest to the most informative one.

\begin{table}[]
\begin{center}
    \caption{Number of runs, among the 100 runs of each requirement-profile combination, in which GP, DT and RS were able to compute a sound assumption }
    \label{tab:resultsForModel}
    \scalebox{0.95}{
    \begin{tabular}{l  r r r    r r r    r r r}
    \toprule
    & \multicolumn{3}{c}{\textbf{IP}} &  \multicolumn{3}{c}{\textbf{IP}$^\prime$} &  \multicolumn{3}{c}{\textbf{IP}$^{\prime\prime}$}\\
    \midrule
 \textbf{Req.}  &  \textbf{GP} & \textbf{DT} & \textbf{RS}  & \textbf{GP} & \textbf{DT} & \textbf{RS} & \textbf{GP} & \textbf{DT} & \textbf{RS} \\
    \midrule
         \texttt{REG-}$\phi_1$  & $79$ &$99$ &$99$ 
           &$76$ & $83$& $100$&$22$ & $0$ & $29$\\
         \texttt{REG-}$\phi_2$  &$29$ &$96$ &$37$ 
          &$21$  &$75$ &$53$ & $11$ &$0$&$17$
          \\
         \texttt{REG-}$\phi_3$  & $76$ &$100$ &$87$ & $30$ &$100$ &$91$ &$3$ & $0$ & $33$ \\
  \texttt{REG-}$\phi_4$  & - & - & - &        $2$ &$18$ &$0$ & $6$ &$1$ &$0$ 
           \\
\texttt{REG-}$\phi_5$ & - & - & - & $43$ &$14$ &$0$ &$11$  &$0$ &$0$ 
         \\
\texttt{REG-}$\phi_6$  & - & - &  -& $1$ &$13$ &$7$ & $9$ &$2$ & $0$ \\
          \texttt{TU-}$\phi_1$ 
          & $42$ &$79$ &$22$ & $52$  &$2$ &$24$ &$8$ & $4$ & $0$
          \\
           \texttt{TU-}$\phi_2$  &$100$ &$94$ &$100$ & $21$ &$96$ &$0$ & $19$ & $0$&$0$ \\
 \texttt{FSM-}$\phi_{1}$  &  - &    -              &- & - & - & -         & 
                  $100$ &$86$&$100$ \\
    \texttt{FSM-}$\phi_{2}$   & - & - & - & - &- &-      & $1$ &$1$ &$0$\\
          \texttt{TT-}$\phi_{1}$  & $96$ &$85$ &$97$ \\
          \texttt{TT-}$\phi_{2}$  &$93$ &$62$ &$89$ \\
          \texttt{TT-}$\phi_{3}$  & $86$ &$59$ &$81$  \\
          \texttt{TT-}$\phi_{4}$ &$93$ &$55$ &$93$\\
          \texttt{TT-}$\phi_{5}$  & $97$ &$54$&$98$\\
          \texttt{TT-}$\phi_{6}$ & $92$ &$55$ &$89$ \\
          \texttt{TT-}$\phi_{7}$  & $84$ &$49$&$79$  \\
          \texttt{TT-}$\phi_{8}$  & $88$ &$94$ &$85$ \\
        \texttt{AC-}$\phi_1$  & $0$ &$0$&$0$  \\
    \bottomrule
    \end{tabular}}
    \end{center}
        $^\ast$ Symbol ``-'' marks entries related with input profiles for which the requirements are violated or satisfied.\\
        For the eight requirements  of Two Tanks~(\texttt{TT}), only the input profile \texttt{IP} was considered since this model only supports constant input signals. \\
        For \texttt{AC}, we considered the input profile \texttt{IP} suggested by our industrial partner.
\end{table}

Considering each of the $32$ combinations of requirements and input profiles separately, 
GP computed a sound assumption for $31$ combinations in at least one of the $100$ runs. 
DT and RS computed, respectively, a sound assumption for $26$ and $22$ of the $32$ combinations in at least one of the~$100$ runs. In the one case that all the three techniques failed to generate a sound assumption, the problem was due to a limitation of QVtrace, which could not terminate the analysis of the assumptions. In the cases where only DT or RS could not compute a sound assumption, the assumption learned by~GP has a complex structure and, therefore, could not be computed by DT and was difficult to be generated by RS. 
 To statistically compare the distributions of the informative values generated by GP with those generated by DT and RS, we used the Wilcoxon rank sum test~\cite{mcdonald2009handbook}  with the level
of significance ($\alpha$) set to $0.05$. 
In both cases, the test rejected the null hypothesis (p-values $<$ $0.05$). Hence,  assumptions learned by GP are significantly more informative than those learned by RS and DT.

 \begin{figure}
\includegraphics[width=\columnwidth]{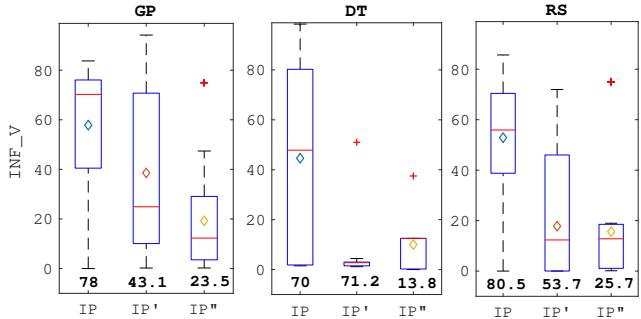}
\caption{
Comparing GP, DT, and RS. The box plots show the informative value of GP, DT, and RS for the input profiles \texttt{IP}, \texttt{IP$^\prime$} and, \texttt{IP$^{\prime\prime}$} (labels on the bottom of the figure). 
 Diamonds depict the average.
The value below the box plot is 
 the percentage of runs, across all the runs of each input profile, in which the technique was able to compute a sound assumption.}
\label{fig:boxplots}
\end{figure}

The box plots in Figure~\ref{fig:boxplots} depict the behavior of GP, DT, and RS across
 the input profiles \texttt{IP}, \texttt{IP$^\prime$} and, \texttt{IP$^{\prime\prime}$}.
 Each box plot 
reports the informative value of one tool for a given input profile and
 is labelled with the percentage of cases, across the runs associated with that input profile, in which the tool was able to compute a sound assumption (value reported below the box plot).
 The results confirm that 
 GP generates  more informative assumptions than
 DT and RS, however, with a negligible loss of capacity to compute sound assumptions.
For GP and DT, the test returned  p-values lower than $0.05$ for \texttt{IP} and \texttt{IP$^\prime$}. 
For \texttt{IP$^{\prime\prime}$},
the p-value is greater than $0.05$ (i.e., $0.1$) since
the sample size is too small to reject the null hypothesis.
For GP and RS, the test returned  p-values lower than $0.05$ for all the input profiles.
 The results  show that, as expected, the more complex the input profile, the more difficult the computation of a sound assumption.
 Therefore, to handle more complex input profiles, developers should tune the values of the parameters in Table~\ref{tab:expGPconfig}, e.g., by increasing the timeout, the number of test cases, and the population size.

 \resq{The answer to \textbf{RQ1} is that, on the considered study subjects, in contrast to DT and RS, GP can learn a sound assumption for $31$ combinations out of $32$ combinations of requirements and input profiles. The assumptions computed by GP are also significantly ($20$\% and $8$\%) more informative  than those learned by DT and RS.
}

\subsection{RQ2-1 --- Usefulness of the Informative Assumptions}
\label{sec:usefulnessv1}
To check whether GP can learn informative assumptions similar to the one manually defined by engineers, we empirically evaluated GP by considering the \texttt{AC} component of the ESAIL microsatellite.
We analyzed the assumptions computed by GP in collaboration with the industrial CPS engineers that developed ESAIL. 
In this question, since we do not limit our analysis to sound assumptions, 
EPIcuRus was configured to return a sound assumption, if found, or the assumption computed in the last iteration of EPIcuRus, if QVtrace was not able to confirm the soundness of any of the assumption generated across the different iterations.

\textbf{Methodology and Experimental Setup.}
To learn assumptions on the 27 input signals of the ten inputs of ESAIL (see Table~\ref{tab:inputs}), 
we considered the parameter settings in Table~\ref{tab:expGPconfig2}. Cells marked with a gray background color denote parameter values that differ from the one considered for RQ1.
We increased the values assigned to the test suite size (\texttt{TS\_Size}) and the timeout (\texttt{Timeout}), since \texttt{AC} is significantly more complex than the other models. Recall from RQ1 that the parameter values in Table~\ref{tab:expGPconfig} did not lead to any sound assumption.
The number of conjunctions (\texttt{Max\_Conj}) and disjunctions  (\texttt{Max\_Disj}) are respectively set to $1$ and $0$ since, according to our industrial partner, 
the assumptions can be represented as a conjunction of two  inequalities among complex arithmetic expressions (see Section~\ref{sec:running}).
In practice, engineers do not know a priori the most informative assumption of the system. 
However, they can select parameter values based on their domain knowledge combined with experiments.
The values assigned to \texttt{Const\_Min} and \texttt{Const\_Max} are, respectively, the lowest and the highest values the input signals of \texttt{AC} can assume.
We considered the input profile \texttt{IP} with a single control point   since, given the time domain (see Section~\ref{sec:running}) and according to the engineers of ESAIL,  
this input profile is sufficiently complex to represent changes in the inputs of \texttt{AC} over the considered time domain.
We assumed the ranges
$[-0.01,0.01]$, 
$[-0.01,0.01]$,
$[0,0.005]$,
$[0,0.005]$, and
$[0,0.005]$
as input domain for each of the input signals of the inputs $\texttt{q}_t$, $\texttt{q}_e$, 
$\omega_t$, 
$\omega_e$, and \texttt{Rwh}, respectively.
We set the other inputs to constant values provided by our industrial partner.

We considered $100$ runs of EPIcuRus and saved the assumptions computed for requirement $\phi_1$ in each of these runs.
Then, to assess whether EPIcuRus can learn assumptions similar to the ones manually defined by engineers, we proceeded as follows.
We elicited the assumption of \texttt{AC} for $\phi_1$ in collaboration with the ESAIL engineers. 
This was done by consulting the Simulink\textsuperscript{\tiny\textregistered} model design and the  design documents of the satellite. 
We then compared the assumptions returned by EPIcuRus and the one we elicited
in collaboration with the ESAIL engineers, i.e., the assumption $\texttt{A}_1$ for  $\phi_1$ (see Section~\ref{sec:running}). 
While eliciting this assumption, we found discrepancies between the model design and the design documents of the satellite. The problems were in the design documents of the satellite and were fixed by ESAIL engineers.
To  assess  the extent to which  GP  can  learn  sound  assumptions  similar to  the  ones  manually  defined  by  engineers,
we analyzed how many runs of EPIcuRus were able to learn each of the terms of the predicates $\texttt{P}_1(t)$ and  $\texttt{P}_2(t)$.
Note that, while the values of \texttt{Const\_Min} and \texttt{Const\_Max} are set to $-0.1$ and $0.1$, coefficients of the terms with higher values (i.e., $+783.3$ in $\texttt{A}_1$) can be generated by selecting small values for the constant terms, i.e., $+1$ and $-1$ in the assumption $\texttt{A}_1$.
Furthermore, given the domain of the inputs, the minimum and the maximum value for each term in $\exp$ is reported in Table~\ref{tab:minmaxvalue}. 
Given the ranges in the table, the term that can assume the highest value in $\exp$ is \texttt{T1}, followed by \texttt{T2}.
For example, for term \texttt{T2} the minimum value is $-1.66$ (i.e., $-332.6\cdot 0.005$) and the maximum value is $0$ (i.e., $-332.6 \cdot 0$).

\begin{table}
\begin{center}
\caption{Values for the parameters of Table~\ref{tab:parameters}  used to assess whether EPIcuRus can learn assumptions similar to the one manually defined by engineers.}
\label{tab:expGPconfig2}
\begin{tabular}{c c  c c  c }
\toprule
& \textbf{Parameter} & \textbf{Value} & \textbf{Parameter} & \textbf{Value} \\
\midrule
\multirow{3}{*}{\textbf{EP}}
& \texttt{SBA} & \texttt{GP} &
\texttt{ST} & $[0, 1]$s\\
& \texttt{TS$\_$Size}  &  \cellcolor{gray!30} $3000$ &
\texttt{Stop$\_$Crt} & \texttt{Timeout} \\
& \texttt{Timeout} &  \cellcolor{gray!30}$5$h
& \texttt{Nbr$\_$Runs} &  $100$\\
\midrule
\multirow{6}{*}{\textbf{GP}}
&
\texttt{Max$\_$Conj} & \cellcolor{gray!30} $1$ 
& \texttt{Max$\_$Disj} & \cellcolor{gray!30} $0$
\\
& \texttt{Const$\_$Min} & \cellcolor{gray!30} $-0.1$ &
\texttt{Const$\_$Max} & \cellcolor{gray!30} $0.1$ \\
&
\texttt{Max$\_$Depth}&  $5$ &
\texttt{Init$\_$Ratio}&  $50\%$ \\ 
& \texttt{Pop$\_$Size} &  $500$ &
\texttt{Gen$\_$Size}&   $100$ \\
&\texttt{Sel$\_$Crt} &  \emph{TRS} &
\texttt{T$\_$Size} & 7\\
& \texttt{Mut$\_$Rate} &  $0.1$  &
 \texttt{Cross$\_$Rate} &  $0.9$ \\
\bottomrule
\end{tabular}
\end{center}
\end{table}

\begin{table}[]
    \caption{Minimum and maximum value of each term of $\exp$.}
    \label{tab:minmaxvalue}
    \centering
    \begin{tabular}{c l l c l l}
        \toprule
        \textbf{Term} & \textbf{Min} & \textbf{Max} &         \textbf{Term} & \textbf{Min} & \textbf{Max} \\
        \midrule
        \texttt{T1} & \phantom{$-$}$0$ & $3.91$ & \texttt{T2} & $-1.66$ &$0$ \\
        \texttt{T3} & \phantom{$-$}$0$ & $0.175$ & \texttt{T4} & $-0.0012$ &$0$ \\
        \texttt{T5} & $-0.1187$ & $0$ & \texttt{T6} & \phantom{$-$}$0$ &$0.0897$ \\
        \texttt{T7} & \phantom{$-$}$0$ & $0.025$ & \texttt{T8} & $-0.025$ & $0$ \\
        \texttt{T9} & $-0.0013$ & $0$ & \texttt{T10} & \phantom{$-$}$0$ &$0.0013$ \\
        \bottomrule
    \end{tabular}
\end{table}

\textbf{Results.} 
We obtained $100$ assumptions from $100$ runs of EPIcuRus. These assumptions could not be proven to be sound by QVtrace due to the complexity of their mathematical expressions. We analyzed and compared the syntax and the semantics of these assumptions with respect to the actual assumption of \texttt{AC} ($\texttt{A}_1$). The results are shown in Table~\ref{tab:resultsP1RQ2}. Specifically, Table~\ref{tab:resultsP1RQ2} shows which of the terms \texttt{T1}, \texttt{T2}, \ldots, \texttt{T10} of $\texttt{P}_1$ and $\texttt{P}_2$, respectively, appear in the $100$ assumptions computed by EPIcuRus.
Each row of the table reports four metrics computed for one of the ten terms of both $\texttt{P}_1$ and $\texttt{P}_2$. 
For each term \texttt{T$_i$} of $\texttt{P}_1$ or $\texttt{P}_2$, the tables' columns show the following: 
\begin{itemize}
\item \texttt{N}-labelled columns. The number of runs  in which the assumptions generated by EPIcuRus contain the term \texttt{T$_i$} but not necessarily with the same coefficient as that appearing in the expression $\exp$. 
\item \texttt{S}-labelled columns. The percentage  of the runs in which  the sign of \texttt{T$_i$} is the same as its sign in the expression $\exp$ over all the runs where the generated assumption by  EPIcuRus contained \texttt{T$_i$}.
\item \texttt{D}-labelled columns. The average of the differences between the values of the coefficients of \texttt{T$_i$} in  $\exp$ and in the assumptions returned by EPIcuRus.
\item \texttt{MaxD}-labelled columns. The maximum of the differences between the values of the coefficients of \texttt{T$_i$} in  $\exp$ and in the assumptions returned by EPIcuRus.
\end{itemize}
The column labeled by \texttt{C} in Table~\ref{tab:resultsP1RQ2} shows the values of the coefficients of the terms \texttt{T$_i$} in $\exp$.

\begin{table}[]
    \caption{The value \texttt{C} of the coefficient of the term \texttt{T$_i$} in $\exp$, The number \texttt{N} of runs in which EPIcuRus was able to learn \texttt{T$_i$} in $\texttt{P}_1$ and $\texttt{P}_2$.
     The percentage \texttt{S} of runs in which the sign of \texttt{T$_i$} was correct. 
    The average \texttt{D} and the maximum \texttt{MaxD} of the difference  between the coefficient of \texttt{T$_i$} of $\texttt{A}_1$ and the one returned by EPIcuRus.
    }
    \label{tab:resultsP1RQ2}
    \centering
    \scalebox{0.87}{
    \begin{tabular}{l |c| r r r r| r r r r}
        \toprule
                &&\multicolumn{4}{c|}{$\textbf{P}_1$}&\multicolumn{4}{c}{$\textbf{P}_2$} \\
        \midrule
\textbf{T$_i$}&\textbf{C}&  \textbf{N}&  \textbf{S}&  \textbf{D}&  \textbf{MaxD} &  \textbf{N} &   \textbf{S}&  \textbf{D}& \textbf{MaxD} \\
  \midrule
 \texttt{T1} &$783$ & $59$& $71$& $679$&$4328$& $88$&$94$ &$256$& $1550$\\
 \texttt{T2}  &$-332$& $69$ & $91$& $247$&$1373$&$74$ &$68$ &$409$&$1331$ \\
 \texttt{T3} & $3$&$3$ & $33$&$21$ &$35$&$9$ &$11$ &$41$&$203$ \\
 \texttt{T4} &$-50$ &$5$ & $40$& $17682$&$87238$&$17$ &$58$ &$30778$& $242066$\\
 \texttt{T5} &$-4751$ &$1$ & $0$& $4751$ &$4751$&$3$ &$33$ &$4949$& $5345$\\
 \texttt{T6} &$3588$ &$1$ & $0$& $3588$ &$3588$&$4$ &$25$ &$3520$&$3998$ \\
 \texttt{T7} &$1000$ &$0$ & - & -&$0$&$0$ &- &- &$0$\\
 \texttt{T8} &$-1000$ &$0$ & - & -&$0$&$0$ &- &- &$0$\\
 \texttt{T9} &$-54$ &$5$ & $60$&$17536$ &$87133$&$6$ &$50$ &$1585$& $4732$\\
 \texttt{T10} &$54$ &$0$ &- &- &$0$&$1$ &$0$ &$54$ &$54$\\
\bottomrule
    \end{tabular}}
\end{table}

The results in  Table~\ref{tab:resultsP1RQ2} show that the terms \texttt{T1} and \texttt{T2} of $\texttt{P}_1$ and $\texttt{P}_2$, that can yield the highest values among other terms (see Table~\ref{tab:minmaxvalue}), were contained in the assumptions returned by EPIcuRus in $59$ and $69$, and $88$ and $74$,
 out of the $100$ runs, respectively.
Note that GP learns assumptions with an arbitrary structure as it does not know the structure of the assumption a priori.
For this reason, the terms \texttt{T1} and \texttt{T2} of $\texttt{P}_1$ are contained in a different number of assumptions than
the terms \texttt{T1} and \texttt{T2} of $\texttt{P}_2$, respectively.
The other terms of $\texttt{A}_1$,
that have  negligible impact on  the value of $\exp$ compared with \texttt{T1} and \texttt{T2}, cannot be effectively learned by EPIcuRus, i.e., they are contained in the assumptions returned by EPIcuRus in only a limited number of runs ($<17$ each).
This is an inherent property of the search as it cannot learn terms that have limited impact on the fitness function.
Note that, in most of the runs ($78$  out of $100$), all the terms learned by EPIcuRus are either part of $\texttt{P}_1$ or part of $\texttt{P}_2$.
Only in $28$ runs the assumptions produced by EPIcuRus contained a spurious term.
Furthermore, in these cases, given the input domains, the values the spurious terms could yield are irrelevant compared to the other terms. To conclude,
since the terms \texttt{T1} and \texttt{T2} of $\texttt{P}_1$ and $\texttt{P}_2$ were contained in the assumptions returned by EPIcuRus in $59$ and $69$, and $88$ and $74$, out of the $100$ runs,  engineers are very likely to learn an assumption that contains the terms \texttt{T1} and \texttt{T2} by executing EPIcuRus a few time, which would take less than a day.
For example, for the term \texttt{T1} of $\texttt{P}_1$ the probability of finding it during the first, second, or third run is $0.9311$ (i.e., $0.59+(1-0.59)\cdot0.59+(1-0.59)^2\cdot0.59$). 
This shows that the term \texttt{T1} of $\texttt{P}_1$ is likely to be computed within three runs.

When the terms \texttt{T1} and \texttt{T2}  were contained in the assumptions computed by EPIcuRus, the sign was correct 
in $71\%$ and $91\%$, and $94\%$ and $68\%$ of the cases, for $\texttt{P}_1$ and $\texttt{P}_2$, respectively.
The average of the differences  between the coefficients of the terms \texttt{T1} and \texttt{T2} of $\texttt{P}_1$ and $\texttt{P}_2$ in $\texttt{A}_1$ and in the assumptions returned by EPIcuRus are $679$ and $247$, and $256$ and $409$, respectively.
The maximum of the differences  between the coefficients of the terms \texttt{T1}, and \texttt{T2} of $\texttt{P}_1$ and $\texttt{P}_2$ in $\texttt{A}_1$ and in the assumptions returned by EPIcuRus are $4328$ and $1373$, and $1550$ and $1331$, respectively.
Note that the coefficients of the terms \texttt{T1}, and \texttt{T2} of $\texttt{P}_1$ and $\texttt{P}_2$ in $\texttt{A}_1$ are $783$ and $-332$, respectively. As we can see, the coefficient values computed by EPIcuRus are not too different from the actual coefficient values for \texttt{T1}, and \texttt{T2}, even though EPIcuRus could technically select any arbitrary number as a coefficient for these terms.  However, provided with such a large search space, EPIcuRus  has been able to select coefficients for these terms that are in the same order of magnitude (i.e., number's nearest power of ten) as their actual  coefficients.  
Such accuracy for coefficient estimates is acceptable when the informativeness of the assumptions is prioritized and what matters is the identification by engineers of the assumptions' terms that have high impact on the satisfaction of the requirement. 
A higher accuracy would not have significant practical benefits since the model checker would anyway not be able to confirm the soundness of the assumption.

To illustrate how useful EPIcuRus can be in practice, let us take three of the $100$ runs for which EPIcuRus returned the following representative assumptions:

\begin{center}
{\footnotesize
\begin{align}
& \texttt{A}_{r1}(t)=  && \underbrace{+1003.4 \cdot \omega_e\_\texttt{x}(t)}_{\texttt{P1-T1}} 
    \underbrace{- 515.8 \cdot \omega_e\_\texttt{y}(t)}_{\texttt{P1-T2}}
    +1 \geq 0\ \wedge \nonumber\\
& && \underbrace{958.0 \cdot \omega_e\_\texttt{x}(t)}_{\texttt{P2-T1}}     
    \underbrace{-452.8  \cdot \omega_e\_\texttt{y}(t)}_{\texttt{P2-T2}}  -1\leq 0 \nonumber\\
 & \texttt{A}_{r2}(t)=  && \underbrace{+757.2 \cdot \omega_e\_\texttt{x}(t)}_{\texttt{P2-T1}} 
    \underbrace{- 413.1 \cdot \omega_e\_\texttt{y}(t)}_{\texttt{P2-T2}} \nonumber  \\
& && \underbrace{-4787.4 \cdot \omega_e\_\texttt{x}(t)\cdot \omega_e\_\texttt{y}(t)}_{\texttt{P2-T4}}\nonumber\\    
& &&  \underbrace{-4787.4  \cdot \omega_e\_\texttt{y}(t)^2}_{\texttt{P2-T9}}  -1\leq 0 \nonumber  \\
& \texttt{A}_{r3}(t)=  && \underbrace{+757.6 \cdot \omega_e\_\texttt{x}(t)}_{\texttt{P1-T1}} 
    \underbrace{- 448.4 \cdot \omega_e\_\texttt{y}(t)}_{\texttt{P1-T2}}\nonumber  \\
& && + 448.4 \cdot \omega_e\_\texttt{x}(t)^2   
    +1 \geq 0\ \wedge \nonumber\\
& && \underbrace{856.8 \cdot \omega_e\_\texttt{x}(t)}_{\texttt{P2-T1}}     
    \underbrace{-419.8  \cdot \omega_e\_\texttt{y}(t)}_{\texttt{P2-T2}} \nonumber \\
    & && - 8.6 \cdot \texttt{Rwh}\_\texttt{z}(t) -1\leq 0 \nonumber  
\end{align}}
\end{center}
where \texttt{P1-T1}, \texttt{P1-T2}, \ldots\ and \texttt{P2-T1}, \texttt{P2-T2}, \ldots\ label the terms \texttt{T1}, \texttt{T2}, \ldots\ of $\texttt{P}_1(t)$ and  $\texttt{P}_2(t)$, respectively.
EPIcuRus returns both assumptions that contain the terms \texttt{T1} and \texttt{T2} of both predicates  (e.g., $\texttt{A}_{r1}(t)$),  and assumptions that contain the terms \texttt{T1} and \texttt{T2} of only one of the predicates (e.g., $\texttt{A}_{r2}(t)$).
Some assumptions contain only terms that are part of $\texttt{A}_1$ (e.g., $\texttt{A}_{r1}(t)$, $\texttt{A}_{r2}(t)$), others contain additional spurious terms (e.g., $\texttt{A}_{r3}(t)$), i.e., terms that are not part of $\texttt{A}_1$ (e.g., $8.6 \cdot \texttt{Rwh}\_\texttt{z}(t)$).
Finally, note that, when EPIcuRus learns 
a term present in both $\texttt{P}_1(t)$ and  $\texttt{P}_2(t)$ (e.g., \texttt{T1}), the coefficients of \texttt{P1-T1} and \texttt{P1-T2} are not necessarily equal.

\textbf{Discussion.} 
When EPIcuRus is configured to learn informative assumptions that are not necessarily sound, our results show that the resulting assumptions include the terms that yield the highest values among other terms in the actual assumption and hence have the most impact on the satisfaction of the requirement under analysis. Even though, in the generated assumptions, not all the terms are present, and the values of their coefficients are only estimates, ESAIL engineers confirmed that these assumptions are still useful and beneficial for designing CPS components, because they can help engineers identify flaws in their components. 

Specifically, engineers generally know which input signals impact the most the satisfaction of a requirement and expect to see those input signal variables in the assumptions generated by EPIcuRus. The absence of those variables in the generated assumptions may indicate flaws. 
For example, by analyzing $\texttt{A}_{r1}(t)$, engineers understand that estimated speeds of the satellite across $\texttt{x}$ and $\texttt{y}$ axes have a significant impact on the satisfaction of $\phi_1$, as expected.
Furthermore, the assumption also provides  high level and approximate information regarding the values of $\omega_e\_\texttt{x}(t)$
and $\omega_e\_\texttt{y}(t)$ that satisfy the requirement.

Engineers can execute several runs of EPIcuRus and obtain a report containing the information shown in Table~\ref{tab:resultsP1RQ2}. 
By consulting the data reported in the table, they can understand which terms are present in most of the assumptions computed by EPIcuRus and therefore have the highest impact on the satisfaction of the requirement under analysis. Running $100$ runs of EPIcuRus requires  approximately $20$ days.
However, the results can be obtained within approximately one day by running $20$ instances of EPIcuRus in parallel. 
This is a reasonable solution for computing assumptions of critical CPS components.

The assumptions returned by EPIcuRus could not be learned by our previous version of EPIcuRus that relies on DT to learn assumptions.
Finally, EPIcuRus is the only existing tool that is able to synthesize assumptions for CPS Simulink\textsuperscript{\tiny\textregistered} components.
Therefore, there is no alternative that engineers could consider to address their need.

\resq{The answer to \textbf{RQ2-1} is that, 
among the terms of the assumptions identified by the LuxSpace engineers, EPIcuRus configured with GP was able to learn the terms that yield the highest values among other terms in the actual assumption.
These assumptions could not be learned with any other tool.
}

\subsection{RQ2-2 --- Usefulness of the Sound Assumptions}
\label{sec:usefulnessv2}
To check whether GP can learn sound assumptions similar to the one manually defined by engineers, we configured EPIcuRus to increase the chances of computing simpler assumptions whose soundness can be verified by QVtrace.

\textbf{Methodology and Experimental Setup.} To check whether learning simpler assumptions allows QVtrace to prove that they are sound, we configured EPIcuRus using the values of the parameters in Table~\ref{tab:expGPconfig2v2}.
Compared with the values in Table~\ref{tab:expGPconfig2}, we decreased the timeout from five hours to one hour, the number of generations (Gen$\_$Size) from $100$ to $10$, and set the values assigned to \texttt{Const\_Min} and \texttt{Const\_Max} to $-0.001$ and $0.001$, respectively.
We considered $100$ runs of EPIcuRus and saved the assumptions computed for the requirement $\phi_1$ in each of these runs. Then, we computed the percentage of the runs in which EPIcuRus was able to compute a sound assumption.

\textbf{Results.} Across the $100$ runs, EPIcuRus was able to compute sound assumptions in $16$ runs. On average, generating one sound assumption for \texttt{AC} took about 6 hours. All the $16$ sound assumptions generated by EPIcuRus were less informative (simpler) than $\texttt{A}_1$, the actual assumption of \texttt{AC}. We identified three distinct patterns to categorize the $16$ generated assumptions. Below, we list the patterns and,  for each one, we show which terms from the original assumption $\texttt{A}_1$ appear in the pattern.

\begin{center}
{\footnotesize
\begin{align}
& \texttt{A}_{s1}(t)=  && \underbrace{\texttt{a} \cdot \omega_e\_\texttt{y}(t)}_{\texttt{P1-T2}} 
    - \texttt{c} \leq 0\ \wedge \nonumber\\
&&&    \underbrace{\texttt{b} \cdot \omega_e\_\texttt{x}(t)}_{\texttt{P2-T1}}  
    - \texttt{d} \leq 0\nonumber\\
& \texttt{A}_{s2}(t)=  && \underbrace{\texttt{a} \cdot \omega_e\_\texttt{x}(t)}_{\texttt{P2-T1}} 
\underbrace{+\texttt{b} \cdot \omega_e\_\texttt{y}(t)}_{\texttt{P2-T2}} 
    - \texttt{c} \leq 0 \nonumber\\
& \texttt{A}_{s3}(t)=  && \underbrace{\texttt{a} \cdot \omega_e\_\texttt{x}(t)}_{\texttt{P2-T1}} 
\underbrace{+\texttt{b} \cdot \omega_e\_\texttt{y}(t)}_{\texttt{P2-T2}} 
\underbrace{+\texttt{c} \cdot \omega_e\_\texttt{z}(t)}_{\texttt{P2-T3}} 
 \leq 0 \nonumber
\end{align}}
\end{center}

The above assumptions, although simpler than the original assumption $\texttt{A}_1$, are still sufficiently informative as confirmed by ESAIL engineers. For example, the $\texttt{A}_{s1}(t)$ pattern indicates that, when the values of the estimated speed of the satellite along its $x$ and $y$ axes are low, requirement $\phi_1$ is satisfied regardless of the values assigned to the other inputs. Similarly, the $\texttt{A}_{s2}(t)$ and $\texttt{A}_{s3}(t)$ patterns indicate that, when the sum of the speeds of the satellite along its $x$ and $y$ axes is low, requirement $\phi_1$ is satisfied.  In other words, despite being an oversimplification of reality, learned assumptions provide correct insights into the conditions leading to the satisfaction of requirements.

\resq{The answer to \textbf{RQ2-2} is that, 
 when EPIcuRus was configured with parameters that lead to the generation of simpler assumptions that could be proven by QVtrace, EPIcuRus was able to generate sound assumptions in $16$ runs out of $100$.  Therefore, through multiple runs, EPIcuRus can generate a sound assumption within approximately six hours. Though simpler than actual assumptions, these learned assumptions appear to provide correct and useful insights. 
}

\begin{table}[]
\caption{Values for the parameters of Table~\ref{tab:parameters}  used for RQ2-2.}
\label{tab:expGPconfig2v2}
\begin{tabular}{c c  c c  c }
\toprule
& \textbf{Parameter} & \textbf{Value} & \textbf{Parameter} & \textbf{Value} \\
\midrule
\multirow{3}{*}{\textbf{EP}}
& \texttt{SBA} & \texttt{GP} &
\texttt{ST} & $[0, 1]$s\\
& \texttt{TS$\_$Size}  & $3000$ &
\texttt{Stop$\_$Crt} & \texttt{Timeout} \\
& \texttt{Timeout} &\cellcolor{gray!30} $1$h
& \texttt{Nbr$\_$Runs} &  $100$\\
\midrule
\multirow{6}{*}{\textbf{GP}}
&
\texttt{Max$\_$Conj} & $1$ 
& \texttt{Max$\_$Disj} & $0$
\\
& \texttt{Const$\_$Min} & \cellcolor{gray!30} $-0.001$ &
\texttt{Const$\_$Max} & \cellcolor{gray!30} $0.001$ \\
&
\texttt{Max$\_$Depth}&  $5$ &
\texttt{Init$\_$Ratio}&  $50\%$ \\ 
& \texttt{Pop$\_$Size} &  $500$ &
\texttt{Gen$\_$Size}&  \cellcolor{gray!30}  $10$ \\
&\texttt{Sel$\_$Crt} &  \emph{TRS} &
\texttt{T$\_$Size} & 7\\
& \texttt{Mut$\_$Rate} &  $0.1$  &
 \texttt{Cross$\_$Rate} &  $0.9$ \\
\bottomrule
\end{tabular}
\end{table}

\subsection{Threats To Validity}
\label{sec:disc}
 The set of  models we selected for the evaluation and their features influence the generalizability of our results. Related to this thread, we note that: First, the public domain benchmark of Simulink models  we used in our study   have been previously used in the literature on testing of CPS models~\cite{NejatiGMBFW19,socrates}; second, the models in the benchmark represent realistic and representative models of CPS components from different domains; third, our industry satellite model represents a realistic and representative CPS model for which we could develop assumptions manually by collaborating with the engineers who had developed those models; fourth,  our results can be further generalized  by additional experiments with diverse types of CPS components and by assessing EPIcuRus over those components.

\section{Related Work}
\label{sec:rl}
This section compares EPIcuRus with the following threads of research:  
(i)~verification, testing and monitoring CPS, 
(ii)~compositional and assume-guarantee reasoning for CPS, 
(iii)~learning assumptions for software components, and 
(iv)~learning the values for unspecified parameters.

\emph{Verification, Testing and Monitoring of CPS.} Approaches to verifying, testing, and monitoring CPS were proposed in the literature (e.g.,~\cite{socrates,Bridging,staliro,menghi2019approximationrefinement,DBLP:conf/cav/FrehseGDCRLRGDM11,DBLP:conf/cav/Tiwari12,ase2020,icse2021,borg2020digital,10.1145/3297280.3297512,7969377,shin2020uncertainty}).
However, these approaches usually assume that the assumptions on the inputs of the CPS component under analysis are already specified.
Those approaches verify, test, and monitor the behavior of the CPS component for the input signals that satisfy those assumptions. 
Our work is complementary to those and, in contrast, it automatically identifies (implicit) assumptions on test inputs. 
Considering those assumptions is an important pre-requisite to ensure that testing and verification results are not overly pessimistic or spurious~\cite{DBLP:conf/tacas/CobleighGP03,Barringer03proofrules,giannakopoulou2004assume,5995279,DBLP:journals/csur/HarmanMZ12,DBLP:journals/iee/GiannakopoulouPB08}. 

\emph{Compositional reasoning.} Assume-guarantee and design by contract approaches were proposed in the literature to support hardware and software verification (e.g.,~\cite{9155827,derler2013cyber,SANGIOVANNIVINCENTELLI2012217,DBLP:journals/fac/MenghiSCG19,DBLP:conf/fase/MenghiSCG18,10.1145/503271.503226,10.1007/BFb0028765,10.1007/3-540-45449-7_11,bernaerts2019validating,10.1145/3377930.3389810}). Assume-guarantee contracts represent the assumptions a CPS component makes about its environment, and the properties it satisfies when these assumptions hold, i.e., its guarantees. Some recent work discusses how to apply assume-guarantee to signal-based modeling formalisms, such as Simulink\textsuperscript{\tiny\textregistered} and analog circuits (e.g.,~\cite{5227075,DBLP:journals/access/NuzzoXOFSMDS14,DBLP:conf/date/NuzzoFIS14}). 
However, these frameworks assume that assumptions and guarantees are manually defined by the designers of the CPS components.
Our work is complementary as the assumptions learned by EPIcuRus can be used within these existing frameworks.
Finally, our work also differs from assume-guarantee testing, where the assumptions defined during software design are used to test the individual components of the system~\cite{DBLP:journals/iee/GiannakopoulouPB08}.

\emph{Learning Assumptions.} 
The problem of automatically inferring assumptions of software components, a.k.a supervisory control problem, 
was widely studied in the literature  (e.g.,~\cite{DBLP:conf/tacas/CobleighGP03,DBLP:books/daglib/0034521,giannakopoulou2002assumption,maoz2019symbolic,Barringer03proofrules,giannakopoulou2004assume,ramadge1987supervisory,ramadge1989control,10.1145/3372020.3391557,piterman2020control}). 
However, the solutions proposed in the literature are solely focused on components specified in finite-state machines and are not applicable to signal-based formalisms (e.g., Simulink\textsuperscript{\tiny\textregistered} models), that are widely used to specify CPS components (see Section~\ref{sec:intro}).

Kampmann et al.~\cite{Kampmann2020} proposed an approach to automatically determine under which circumstances a particular program behavior, such as a failure, takes place. 
However, this approach uses a decision tree learner  to observe and learn which input features
are associated with the particular program behavior under analysis. 
As such, for our usage scenario, this approach is going to inherit the same limitations of our earlier version of EPIcuRus.

Dynamic invariants generators (e.g.,~\cite{ERNST200735}) infer conditions that hold at certain points of a program. 
They generate a set of candidate invariants and return the best candidates that hold over the observed program executions.
However, they have not been applied to Simulink\textsuperscript{\tiny\textregistered} models.

Property inference aims at automatically detecting properties that hold in a given system. 
It was also recently applied to feed-forward neural network~\cite{Gopinath2019}.
While many approaches for property inference were proposed in the literature (e.g.,~\cite{houdini, daikon2007, LMN14, Garg2016}), they do not consider signal-based modeling formalisms (e.g., Simulink\textsuperscript{\tiny\textregistered}) which are the targets of this work.

Template-based specification mining are used to synthesize assumptions following with a certain structure~\cite{DBLP:conf/memocode/LiDS11,DBLP:conf/fmcad/AlurMT13}.
However, solutions from the literature (e.g.,~\cite{DBLP:conf/memocode/LiDS11,DBLP:conf/fmcad/AlurMT13,keegan2020control}) use  LTL-GR(1) to express assumptions.
These formalisms are substantially different and less expressive than the one considered in this work, and can not express the signal-based assumptions generated by EPIcuRus.

In this work, we combined model checking and model testing to learn assumptions.
This idea was supported by a recent study~\cite{NejatiGMBFW19} that analyzed the complementarity between model testing and model checking for fault detection purposes. 

Finally, Sato et al.~\cite{sato2020constrained} recently considered the problem of finding an input signal under which the system’s behavior satisfies a
given requirement. 
This is a sub-problem of the assumption generation problem, where assumptions identify conditions on the input signals (and therefore sets of input signals) that ensure the satisfaction of a given requirement.

This paper significantly extends our previous version of EPIcuRus~\cite{epicurus20}. 
Our extension enables learning assumptions containing conditions defined over multiple signals related by both arithmetic and relational operators. 
Assumptions containing conditions defined over multiple signals related by both arithmetic and relational operators are common for industrial CPS components.
This is confirmed by our industrial case study from the microsatellite domain.
Differently than our previous work, we used genetic programming to synthesize complex assumptions of CPS components.
Finally, we performed an extensive and thorough evaluation of the assumptions computed by the extended version of EPIcuRus using an industrial case study from the microsatellite domain. 
The assumptions computed by EPIcuRus were evaluated in collaboration with the engineers that developed the microsatellite.

\emph{Learning Parameters.} The problem of learning (requirement) parameters from simulations was extensively studied in the literature~\cite{DBLP:journals/tcad/JinDDS15,DBLP:conf/hybrid/JinDDS13,6032536}. 
Our work is significantly different from those, since it aims to learn assumptions on the input signals. 

EPIcuRus extends counterexample-guided inductive synthesis~\cite{DBLP:conf/asplos/Solar-LezamaTBSS06} by exhaustively veryfing the learned assumptions using an SMT-based model checker. 
Furthermore, differently from counterexample-guided inductive synthesis, EPIcuRus 
(i)~targets signal-based formalisms, that are widely used in the CPS  industry, (ii)~extracts assumptions from test data, and 
(iii)~uses test cases to efficiently produce a large amount of data to be fed in our machine learning algorithm to derive informative and sound assumptions.
 
\section{Conclusion}
\label{sec:conclusion}
 This paper proposes a technique to learn complex assumptions for CPS systems and components. Our technique uses  genetic programming (GP) to learn assumptions containing conditions defined over multiple signals related by both arithmetic and relational operators. 
Environment  assumptions are required to ensure that the CPS under analysis meets its requirements and to avoid spurious failures during verification.   We evaluated our approach using $12$  models of CPS components with $94$ requirements provided by Lockheed Martin~\cite{lockheedmartin}
and the model of the attitude control component of a microsatellite with four requirements provided by LuxSpace~\cite{luxspace}. 
Our evaluation shows that our approach can learn many more sound environment assumptions compared to the  alternative baseline techniques. Further, our approach is able to learn assumptions that are significantly more informative than those generated by existing techniques. Finally, for our industrial CPS model, our approach is able to generate assumptions that are sufficiently close to the assumptions manually developed by engineers to be of practical value.

\section*{Acknowledgement}
The experiments presented in this paper were carried out
using the HPC facilities of the University of Luxembourg~\cite{VBCG_HPCS14} (\href{http://hpc.uni.lu}{HPC @ Uni.lu}).
 \IEEEcompsocitemizethanks{This project has received funding from the Luxembourg National Research Fund under the grant BRIDGES18/IS/12632261, the European Research Council (ERC) under the European Union's Horizon 2020 research and innovation programme (grant agreement No 694277), and NSERC of Canada under the Discovery and CRC programs.}

\ifCLASSOPTIONcaptionsoff
  \newpage
\fi

\bibliographystyle{IEEEtran}

\begin{thebibliography}{10}
\providecommand{\url}[1]{#1}
\csname url@samestyle\endcsname
\providecommand{\newblock}{\relax}
\providecommand{\bibinfo}[2]{#2}
\providecommand{\BIBentrySTDinterwordspacing}{\spaceskip=0pt\relax}
\providecommand{\BIBentryALTinterwordstretchfactor}{4}
\providecommand{\BIBentryALTinterwordspacing}{\spaceskip=\fontdimen2\font plus
\BIBentryALTinterwordstretchfactor\fontdimen3\font minus
  \fontdimen4\font\relax}
\providecommand{\BIBforeignlanguage}[2]{{
\expandafter\ifx\csname l@#1\endcsname\relax
\typeout{** WARNING: IEEEtran.bst: No hyphenation pattern has been}
\typeout{** loaded for the language `#1'. Using the pattern for}
\typeout{** the default language instead.}
\else
\language=\csname l@#1\endcsname
\fi
#2}}
\providecommand{\BIBdecl}{\relax}
\BIBdecl

\bibitem{giannakopoulou2002assumption}
D.~Giannakopoulou, C.~S. Pasareanu, and H.~Barringer, ``Assumption generation
  for software component verification,'' in \emph{International Conference on
  Automated Software Engineering}.\hskip 1em plus 0.5em minus 0.4em\relax IEEE,
  2002, pp. 3--12.

\bibitem{10.1145/3270112.3270115}
A.~Schaap, G.~Marks, V.~Pantelic, M.~Lawford, G.~Selim, A.~Wassyng, and
  L.~Patcas, ``Documenting simulink designs of embedded systems,'' in
  \emph{International Conference on Model Driven Engineering Languages and
  Systems ({MODELS}): Companion Proceedings}.\hskip 1em plus 0.5em minus
  0.4em\relax ACM, 2018, p. 47–51.

\bibitem{derler2013cyber}
P.~Derler, E.~A. Lee, S.~Tripakis, and M.~T{\"o}rngren, ``Cyber-physical system
  design contracts,'' in \emph{International Conference on Cyber-Physical
  Systems}.\hskip 1em plus 0.5em minus 0.4em\relax ACM, 2013, pp. 109--118.

\bibitem{SANGIOVANNIVINCENTELLI2012217}
``Taming dr. frankenstein: Contract-based design for cyber-physical systems*,''
  vol.~18, no.~3, 2012, pp. 217 -- 238.

\bibitem{10.1007/BFb0028765}
T.~A. Henzinger, S.~Qadeer, and S.~K. Rajamani, ``You assume, we guarantee:
  Methodology and case studies,'' in \emph{Computer Aided Verification}.\hskip
  1em plus 0.5em minus 0.4em\relax Springer, 1998, pp. 440--451.

\bibitem{10.1007/3-540-45449-7_11}
L.~de~Alfaro and T.~A. Henzinger, ``Interface theories for component-based
  design,'' in \emph{Embedded Software}, T.~A. Henzinger and C.~M. Kirsch,
  Eds.\hskip 1em plus 0.5em minus 0.4em\relax Springer, 2001, pp. 148--165.

\bibitem{10.1145/3372020.3391557}
D.~G. Cavezza, D.~Alrajeh, and A.~Gy\"{o}rgy, ``Minimal assumptions refinement
  for realizable specifications,'' in \emph{International Conference on Formal
  Methods in Software Engineering (FormaliSE)}.\hskip 1em plus 0.5em minus
  0.4em\relax ACM, 2020.

\bibitem{giannakopoulou2004assume}
D.~Giannakopoulou, C.~S. Pasareanu, and J.~M. Cobleigh, ``Assume-guarantee
  verification of source code with design-level assumptions,'' in
  \emph{International Conference on Software Engineering}.\hskip 1em plus 0.5em
  minus 0.4em\relax IEEE, 2004, pp. 211--220.

\bibitem{DBLP:conf/tacas/CobleighGP03}
J.~M. Cobleigh, D.~Giannakopoulou, and C.~S. Pasareanu, ``Learning assumptions
  for compositional verification,'' in \emph{Tools and Algorithms for the
  Construction and Analysis of Systems ({TACAS})}.\hskip 1em plus 0.5em minus
  0.4em\relax Springer, 2003, pp. 331--346.

\bibitem{epicurus20}
K.~Gaaloul, C.~Menghi, S.~Nejati, L.~C. Briand, and D.~Wolfe, ``Mining
  assumptions for software components using machine learning,'' in
  \emph{Foundations of Software Engineering ({ESEC/FSE})}.\hskip 1em plus 0.5em
  minus 0.4em\relax ACM, 2020.

\bibitem{fitctree}
\BIBentryALTinterwordspacing
(2020) fitctree. [Online]. Available:
  \url{https://nl.mathworks.com/help/stats/fitctree.html}
\BIBentrySTDinterwordspacing

\bibitem{koza1992genetic}
J.~R. Koza and J.~R. Koza, \emph{Genetic programming: on the programming of
  computers by means of natural selection}.\hskip 1em plus 0.5em minus
  0.4em\relax MIT press, 1992, vol.~1.

\bibitem{poli2008field}
R.~Poli, W.~B. Langdon, N.~F. McPhee, and J.~R. Koza, \emph{A field guide to
  genetic programming}.\hskip 1em plus 0.5em minus 0.4em\relax Lulu. com, 2008.

\bibitem{banzhaf1998genetic}
W.~Banzhaf, P.~Nordin, R.~E. Keller, and F.~D. Francone, \emph{Genetic
  programming}.\hskip 1em plus 0.5em minus 0.4em\relax Springer, 1998.

\bibitem{GPMatlab}
\BIBentryALTinterwordspacing
(2020) Matlab {GP} toolbox. [Online]. Available:
  \url{https://it.mathworks.com/matlabcentral/fileexchange/47197-genetic-programming-matlab-toolbox}
\BIBentrySTDinterwordspacing

\bibitem{madar2005genetic}
J.~Mad{\'a}r, J.~Abonyi, and F.~Szeifert, ``Genetic programming for the
  identification of nonlinear input- output models,'' \emph{Industrial \&
  engineering chemistry research}, vol.~44, no.~9, pp. 3178--3186, 2005.

\bibitem{montana1995strongly}
D.~J. Montana, ``Strongly typed genetic programming,'' \emph{Evolutionary
  computation}, vol.~3, no.~2, pp. 199--230, 1995.

\bibitem{lockheedmartin}
\BIBentryALTinterwordspacing
(2020) {Lockheed Martin}. [Online]. Available:
  \url{https://www.lockheedmartin.com/en-us/index.html}
\BIBentrySTDinterwordspacing

\bibitem{luxspace}
\BIBentryALTinterwordspacing
``Luxspace,'' 2020. [Online]. Available: \url{https://luxspace.lu/}
\BIBentrySTDinterwordspacing

\bibitem{esa}
\BIBentryALTinterwordspacing
``{The European Space Agency} ({ESA}),'' 2020. [Online]. Available:
  \url{https://www.esa.int/}
\BIBentrySTDinterwordspacing

\bibitem{exactEarth}
\BIBentryALTinterwordspacing
``{exactEarth},'' 2020. [Online]. Available: \url{https://www.exactearth.com/}
\BIBentrySTDinterwordspacing

\bibitem{phases}
\BIBentryALTinterwordspacing
{ESA}, ``Building and testing spacecraft,'' 2020. [Online]. Available:
  \url{https://www.esa.int/Science_Exploration/Space_Science/Building_and_testing_spacecraft}
\BIBentrySTDinterwordspacing

\bibitem{mathworks}
``{Mathworks},'' \url{https://mathworks.com}, 02 2019.

\bibitem{menghi2019approximationrefinement}
C.~Menghi, S.~Nejati, L.~C. Briand, and Y.~I. Parache,
  ``Approximation-refinement testing of compute-intensive cyber-physical
  models: An approach based on system identification,'' in \emph{International
  Conference on Software Engineering ({ICSE})}.\hskip 1em plus 0.5em minus
  0.4em\relax {ACM}, 2020.

\bibitem{subsystem}
\BIBentryALTinterwordspacing
(2020) Subsystems. [Online]. Available:
  \url{https://it.mathworks.com/help/simulink/subsystems.html}
\BIBentrySTDinterwordspacing

\bibitem{sfunction}
\BIBentryALTinterwordspacing
(2020) {S}-{F}unction. [Online]. Available:
  \url{https://www.mathworks.com/help/simulink/sfg/what-is-an-s-function.html}
\BIBentrySTDinterwordspacing

\bibitem{mex}
\BIBentryALTinterwordspacing
(2020) {MEX} function. [Online]. Available:
  \url{https://it.mathworks.com/help/matlab/call-mex-file-functions.html}
\BIBentrySTDinterwordspacing

\bibitem{QVtrace}
\BIBentryALTinterwordspacing
(2020) {QVtrace}. [Online]. Available: \url{https://qracorp.com/qvtrace/}
\BIBentrySTDinterwordspacing

\bibitem{qracorp}
\BIBentryALTinterwordspacing
(2019) {QRA} corp. [Online]. Available: \url{https://qracorp.com/}
\BIBentrySTDinterwordspacing

\bibitem{wie1998space}
B.~Wie, \emph{Space Vehicle Dynamics and Control}, ser. AIAA education
  series.\hskip 1em plus 0.5em minus 0.4em\relax American Institute of
  Aeronautics and Astronautics, 1998.

\bibitem{virtualVectors}
\BIBentryALTinterwordspacing
(2020) Virtual vector. [Online]. Available:
  \url{https://it.mathworks.com/help/simulink/ug/signal-types.html}
\BIBentrySTDinterwordspacing

\bibitem{acquatella2018fast}
P.~Acquatella, ``Fast slew maneuvers for the high-torque-wheels biros
  satellite,'' \emph{Transactions of the Japan Society for Aeronautical and
  Space Sciences}, vol.~61, no.~2, pp. 79--86, 2018.

\bibitem{Chaturvedi2009MSS1823037}
D.~K. Chaturvedi, \emph{Modeling and Simulation of Systems Using MATLAB and
  Simulink}, 1st~ed.\hskip 1em plus 0.5em minus 0.4em\relax Boca Raton, FL,
  USA: CRC Press, Inc., 2009.

\bibitem{Simulink}
\BIBentryALTinterwordspacing
(2020) {Simulink}. [Online]. Available:
  \url{https://nl.mathworks.com/products/simulink.html}
\BIBentrySTDinterwordspacing

\bibitem{de2008z3}
L.~De~Moura and N.~Bj{\o}rner, ``{Z3}: An efficient {SMT} solver,'' in
  \emph{International conference on Tools and Algorithms for the Construction
  and Analysis of Systems}.\hskip 1em plus 0.5em minus 0.4em\relax Springer,
  2008, pp. 337--340.

\bibitem{socrates}
C.~Menghi, S.~Nejati, K.~Gaaloul, and L.~C. Briand, ``Generating automated and
  online test oracles for simulink models with continuous and uncertain
  behaviors,'' in \emph{Foundations of Software Engineering ({ESEC/SIGSOFT}
  {FSE})}.\hskip 1em plus 0.5em minus 0.4em\relax {ACM}, 2019.

\bibitem{staliro}
Y.~Annpureddy, C.~Liu, G.~Fainekos, and S.~Sankaranarayanan, ``{S-TaLiRo}: A
  tool for temporal logic falsification for hybrid systems,'' in \emph{Tools
  and Algorithms for the Construction and Analysis of Systems}.\hskip 1em plus
  0.5em minus 0.4em\relax Springer, 2011, pp. 254--257.

\bibitem{survey}
R.~Huang, W.~Sun, Y.~Xu, H.~Chen, D.~Towey, and X.~Xia, ``A survey on adaptive
  random testing.''\hskip 1em plus 0.5em minus 0.4em\relax IEEE Transactions on
  Software Engineering, 2019.

\bibitem{DBLP:conf/issta/ArcuriB11}
A.~Arcuri and L.~C. Briand, ``Adaptive random testing: an illusion of
  effectiveness?'' in \emph{International Symposium on Software Testing and
  Analysis, {ISSTA}}.\hskip 1em plus 0.5em minus 0.4em\relax {ACM}, 2011, pp.
  265--275.

\bibitem{DBLP:conf/asian/ChenLM04}
T.~Y. Chen, H.~Leung, and I.~K. Mak, ``Adaptive random testing,'' in
  \emph{Advances in Computer Science - {ASIAN}}.\hskip 1em plus 0.5em minus
  0.4em\relax Springer, 2004, pp. 320--329.

\bibitem{signals19}
C.~E. Tuncali, G.~Fainekos, D.~Prokhorov, H.~Ito, and J.~Kapinski,
  ``Requirements-driven test generation for autonomous vehicles with machine
  learning components,'' \emph{IEEE Transactions on Intelligent Vehicles},
  vol.~5, no.~2, pp. 265--280, 2020.

\bibitem{pareto18}
A.~Arrieta, S.~Wang, U.~Markiegi, A.~Arruabarrena, L.~Etxeberria, and
  G.~Sagardui, ``Pareto efficient multi-objective black-box test case selection
  for simulation-based testing,'' \emph{Information and Software Technology},
  vol. 114, pp. 137--154, 2019.

\bibitem{floor}
\BIBentryALTinterwordspacing
(2020) {floor}. [Online]. Available:
  \url{https://www.mathworks.com/help/matlab/ref/floor.html}
\BIBentrySTDinterwordspacing

\bibitem{luke2013essentials}
M.~A. Lones, ``Sean luke: essentials of metaheuristics,'' \emph{Genet. Program.
  Evolvable Mach.}, vol.~12, no.~3, pp. 333--334, 2011.

\bibitem{onepointcrossover}
R.~Poli and W.~B. Langdon, ``Genetic programming with one-point crossover,'' in
  \emph{Soft Computing in Engineering Design and Manufacturing}.\hskip 1em plus
  0.5em minus 0.4em\relax Springer, 1998, pp. 180--189.

\bibitem{poli1998schema}
R.~Poli and W.~B.~Langdon, ``Schema theory for genetic programming with
  one-point crossover and point mutation,'' in \emph{Evolutionary Computation},
  1998, vol.~6, no.~3, pp. 231--252.

\bibitem{9218211}
A.~{Mavridou}, H.~{Bourbouh}, D.~{Giannakopoulou}, T.~{Pressburger},
  M.~{Hejase}, P.~L. {Garoche}, and J.~{Schumann}, ``{The Ten Lockheed Martin
  Cyber-Physical Challenges: Formalized, Analyzed, and Explained},'' in
  \emph{International Requirements Engineering Conference ({RE})}.\hskip 1em
  plus 0.5em minus 0.4em\relax {IEEE}, 2020, pp. 300--310.

\bibitem{frank2016weka}
E.~Frank, M.~A. Hall, and I.~H. Witten, \emph{The WEKA workbench}.\hskip 1em
  plus 0.5em minus 0.4em\relax Morgan Kaufmann, 2016.

\bibitem{silva2003gplab}
S.~Silva and J.~Almeida, ``Gplab-a genetic programming toolbox for {Matlab},''
  in \emph{Proceedings of the Nordic MATLAB conference}.\hskip 1em plus 0.5em
  minus 0.4em\relax Citeseer, 2003, pp. 273--278.

\bibitem{searson2015gptips}
D.~P. Searson, ``{GPTIPS 2}: an open-source software platform for symbolic data
  mining,'' in \emph{Handbook of genetic programming applications}.\hskip 1em
  plus 0.5em minus 0.4em\relax Springer, 2015, pp. 551--573.

\bibitem{githubmaterial}
\BIBentryALTinterwordspacing
(2020) Additional material. [Online]. Available:
  \url{https://github.com/SNTSVV/EPIcuRus}
\BIBentrySTDinterwordspacing

\bibitem{NejatiGMBFW19}
S.~Nejati, K.~Gaaloul, C.~Menghi, L.~C. Briand, S.~Foster, and D.~Wolfe,
  ``Evaluating model testing and model checking for finding requirements
  violations in simulink models,'' in \emph{Foundations of Software
  Engineering}, ser. ESEC/FSE.\hskip 1em plus 0.5em minus 0.4em\relax ACM,
  2019.

\bibitem{arcuri2013parameter}
A.~Arcuri and G.~Fraser, ``Parameter tuning or default values? an empirical
  investigation in search-based software engineering,'' \emph{Empirical
  Software Engineering}, vol.~18, no.~3, pp. 594--623, 2013.

\bibitem{ernst2019arch}
G.~Ernst, P.~Arcaini, A.~Donze, G.~Fainekos, L.~Mathesen, G.~Pedrielli,
  S.~Yaghoubi, Y.~Yamagata, and Z.~Zhang, ``{ARCH-COMP} 2019 category report:
  Falsification.'' in \emph{{ARCH@ CPSIoTWeek}}, 2019, pp. 129--140.

\bibitem{ARCH20}
G.~Ernst, P.~Arcaini, I.~Bennani, A.~Donze, G.~Fainekos, G.~Frehse,
  L.~Mathesen, C.~Menghi, G.~Pedrielli, M.~Pouzet, S.~Yaghoubi, Y.~Yamagata,
  and Z.~Zhang, ``{ARCH-COMP} 2020 category report: Falsification,'' in
  \emph{International Workshop on Applied Verification of Continuous and Hybrid
  Systems ({ARCH20})}, ser. EPiC Series in Computing, vol.~74.\hskip 1em plus
  0.5em minus 0.4em\relax EasyChair, 2020, pp. 140--152.

\bibitem{VBCG_HPCS14}
S.~Varrette, P.~Bouvry, H.~Cartiaux, and F.~Georgatos, ``{Management of an
  Academic HPC Cluster: The UL Experience},'' in \emph{Proc. of the 2014 Intl.
  Conf. on High Performance Computing \& Simulation (HPCS 2014)}.\hskip 1em
  plus 0.5em minus 0.4em\relax Bologna, Italy: IEEE, July 2014, pp. 959--967.

\bibitem{mcdonald2009handbook}
J.~H. McDonald, \emph{Handbook of biological statistics}, 2009, vol.~2.

\bibitem{Bridging}
A.~Mavridou, H.~Bourbouh, P.-L. Garoche, D.~Giannakopoulou, T.~Pressburger, and
  J.~Schumann, ``Bridging the gap between requirements and simulink model
  analysis,'' in \emph{Requirements Engineering: Foundation for Software
  Quality ({REFSQ}), Companion Proceedings}.\hskip 1em plus 0.5em minus
  0.4em\relax {Springer}, 2020.

\bibitem{DBLP:conf/cav/FrehseGDCRLRGDM11}
G.~Frehse, C.~L. Guernic, A.~Donz{\'{e}}, S.~Cotton, R.~Ray, O.~Lebeltel,
  R.~Ripado, A.~Girard, T.~Dang, and O.~Maler, ``{SpaceEx: Scalable
  Verification of Hybrid Systems},'' in \emph{Computer Aided Verification
  ({CAV})}.\hskip 1em plus 0.5em minus 0.4em\relax Springer, 2011.

\bibitem{DBLP:conf/cav/Tiwari12}
A.~Tiwari, ``Hybridsal relational abstracter,'' in \emph{Computer Aided
  Verification ({CAV})}.\hskip 1em plus 0.5em minus 0.4em\relax Springer, 2012.

\bibitem{ase2020}
C.~Boufaied, C.~Menghi, D.~Bianculli, L.~Briand, and Y.~Isasi-Parache,
  ``Trace-checking signal-based temporal properties: A model-driven approach,''
  in \emph{International Conference on Automated Software Engineering (ASE
  2020)}.\hskip 1em plus 0.5em minus 0.4em\relax IEEE, 2020.

\bibitem{icse2021}
C.~Menghi, E.~Viganò, D.~Bianculli, and L.~C. Briand, ``{Trace-Checking CPS
  Properties: Bridging the Cyber-Physical Gap},'' in \emph{International
  Conference on Software Engineering (ICSE)}.\hskip 1em plus 0.5em minus
  0.4em\relax {ACM}, 2021.

\bibitem{borg2020digital}
M.~Borg, R.~B. Abdessalem, S.~Nejati, F.-X. Jegeden, and D.~Shin, ``Digital
  twins are not monozygotic--cross-replicating adas testing in two
  industry-grade automotive simulators,'' \emph{arXiv preprint
  arXiv:2012.06822}, 2020.

\bibitem{10.1145/3297280.3297512}
U.~Markiegi, A.~Arrieta, L.~Etxeberria, and G.~Sagardui, ``Test case selection
  using structural coverage in software product lines for time-budget
  constrained scenarios,'' in \emph{SIGAPP Symposium on Applied Computin
  (SAC)}.\hskip 1em plus 0.5em minus 0.4em\relax ACM, 2019, p. 2362–2371.

\bibitem{7969377}
A.~{Arrieta}, S.~{Wang}, U.~{Markiegi}, G.~{Sagardui}, and L.~{Etxeberria},
  ``Search-based test case generation for cyber-physical systems,'' in
  \emph{IEEE Congress on Evolutionary Computation (CEC)}.\hskip 1em plus 0.5em
  minus 0.4em\relax IEEE, 2017, pp. 688--697.

\bibitem{shin2020uncertainty}
S.~Y. Shin, K.~Chaouch, S.~Nejati, M.~Sabetzadeh, L.~C. Briand, and F.~Zimmer,
  ``Uncertainty-aware specification and analysis for hardware-in-the-loop
  testing of cyber-physical systems,'' \emph{Journal of Systems and Software},
  vol. 171, p. 110813, 2020.

\bibitem{Barringer03proofrules}
H.~Barringer and D.~Giannakopoulou, ``Proof rules for automated compositional
  verification through learning,'' in \emph{In Proc. SAVCBS Workshop}, 2003,
  pp. 14--21.

\bibitem{5995279}
P.~{Derler}, E.~A. {Lee}, and A.~{Sangiovanni Vincentelli}, ``Modeling
  cyber-physical systems,'' \emph{Proceedings of the IEEE}, vol. 100, no.~1,
  pp. 13--28, 2012.

\bibitem{DBLP:journals/csur/HarmanMZ12}
M.~Harman, S.~A. Mansouri, and Y.~Zhang, ``Search-based software engineering:
  Trends, techniques and applications,'' \emph{ACM Computing Surveys ({CSUR})},
  vol.~45, no.~1, pp. 11:1--11:61, 2012.

\bibitem{DBLP:journals/iee/GiannakopoulouPB08}
D.~Giannakopoulou, C.~S. Pasareanu, and C.~Blundell, ``Assume-guarantee testing
  for software components,'' \emph{{IET} Software}, vol.~2, no.~6, pp.
  547--562, 2008.

\bibitem{9155827}
A.~{Arrieta}, J.~A. {Agirre}, and G.~{Sagardui}, ``A tool for the automatic
  generation of test cases and oracles for simulation models based on
  functional requirements,'' in \emph{Software Testing, Verification and
  Validation Workshops (ICSTW)}.\hskip 1em plus 0.5em minus 0.4em\relax {IEEE},
  2020, pp. 1--5.

\bibitem{DBLP:journals/fac/MenghiSCG19}
C.~Menghi, P.~Spoletini, M.~Chechik, and C.~Ghezzi, ``A verification-driven
  framework for iterative design of controllers,'' \emph{Formal Aspects of
  Computing}, vol.~31, no.~5, pp. 459--502, 2019.

\bibitem{DBLP:conf/fase/MenghiSCG18}
C.~Menghi, P.~Spoletini, M.~Checkik, and G.~Carlo, ``Supporting
  verification-driven incremental distributed design of components,'' in
  \emph{Fundamental Approaches to Software Engineering ({FASE})}.\hskip 1em
  plus 0.5em minus 0.4em\relax Springer, 2018.

\bibitem{10.1145/503271.503226}
\BIBentryALTinterwordspacing
L.~de~Alfaro and T.~A. Henzinger, ``Interface automata,'' \emph{ACM SIGSOFT
  Software Engineering Notes}, vol.~26, no.~5, Sep. 2001. [Online]. Available:
  \url{https://doi.org/10.1145/503271.503226}
\BIBentrySTDinterwordspacing

\bibitem{bernaerts2019validating}
M.~Bernaerts, B.~Oakes, K.~Vanherpen, B.~Aelvoet, H.~Vangheluwe, and J.~Denil,
  ``Validating industrial requirements with a contract-based approach,'' in
  \emph{International Conference on Model Driven Engineering Languages and
  Systems Companion (MODELS-C)}.\hskip 1em plus 0.5em minus 0.4em\relax IEEE,
  2019, pp. 18--27.

\bibitem{10.1145/3377930.3389810}
A.~Arrieta, J.~A. Agirre, and G.~Sagardui, ``Seeding strategies for
  multi-objective test case selection: An application on simulation-based
  testing,'' in \emph{Genetic and Evolutionary Computation Conference
  ({GECCO})}.\hskip 1em plus 0.5em minus 0.4em\relax ACM, 2020, p. 1222–1231.

\bibitem{5227075}
X.~{Sun}, P.~{Nuzzo}, C.~{Wu}, and A.~{Sangiovanni-Vincentelli},
  ``Contract-based system-level composition of analog circuits,'' in
  \emph{Design Automation Conference}.\hskip 1em plus 0.5em minus 0.4em\relax
  ACM, 2009.

\bibitem{DBLP:journals/access/NuzzoXOFSMDS14}
P.~Nuzzo, H.~Xu, N.~Ozay, J.~B. Finn, A.~L. Sangiovanni{-}Vincentelli, R.~M.
  Murray, A.~Donz{\'{e}}, and S.~A. Seshia, ``A contract-based methodology for
  aircraft electric power system design,'' \emph{{IEEE} Access}, vol.~2, pp.
  1--25, 2014.

\bibitem{DBLP:conf/date/NuzzoFIS14}
P.~Nuzzo, J.~B. Finn, A.~Iannopollo, and A.~L. Sangiovanni{-}Vincentelli,
  ``Contract-based design of control protocols for safety-critical
  cyber-physical systems,'' in \emph{Design, Automation {\&} Test in Europe
  Conference {\&} Exhibition, ({DATE})}.\hskip 1em plus 0.5em minus 0.4em\relax
  European Design and Automation Association, 2014.

\bibitem{DBLP:books/daglib/0034521}
C.~G. Cassandras and S.~Lafortune, \emph{Introduction to Discrete Event
  Systems, Second Edition}.\hskip 1em plus 0.5em minus 0.4em\relax Springer,
  2008.

\bibitem{maoz2019symbolic}
S.~Maoz, J.~O. Ringert, and R.~Shalom, ``Symbolic repairs for {GR}(1)
  specifications,'' in \emph{International Conference on Software Engineering
  (ICSE)}.\hskip 1em plus 0.5em minus 0.4em\relax IEEE, 2019, pp. 1016--1026.

\bibitem{ramadge1987supervisory}
P.~J. Ramadge and W.~M. Wonham, ``Supervisory control of a class of discrete
  event processes,'' \emph{SIAM journal on control and optimization}, vol.~25,
  no.~1, pp. 206--230, 1987.

\bibitem{ramadge1989control}
P.~J. Ramadge and M.~Wonham~W, ``The control of discrete event systems,''
  \emph{Proceedings of the IEEE}, vol.~77, no.~1, pp. 81--98, 1989.

\bibitem{piterman2020control}
N.~Piterman, M.~Keegan, V.~Braberman, N.~D'Ippolito, and S.~Uchitel, ``Control
  and discovery of reactive system environments,'' in \emph{University of
  Leicester}, 2020.

\bibitem{Kampmann2020}
A.~Kampmann, N.~Havrikov, E.~O. Soremekun, and A.~Zeller, ``When does my
  program do this? learning circumstances of software behavior,'' in
  \emph{Foundations of Software Engineering ({ESEC/FSE})}.\hskip 1em plus 0.5em
  minus 0.4em\relax ACM, 2020.

\bibitem{ERNST200735}
M.~D. Ernst, J.~H. Perkins, P.~J. Guo, S.~McCamant, C.~Pacheco, M.~S. Tschantz,
  and C.~Xiao, ``{The Daikon system for dynamic detection of likely
  invariants},'' \emph{Science of Computer Programming}, vol.~69, no.~1, pp. 35
  -- 45, 2007.

\bibitem{Gopinath2019}
D.~Gopinath, H.~Converse, C.~S. P\u{a}s\u{a}reanu, and A.~Taly, ``Property
  inference for deep neural networks,'' in \emph{International Conference on
  Automated Software Engineering ({ASE})}.\hskip 1em plus 0.5em minus
  0.4em\relax IEEE, 2019, p. 797–809.

\bibitem{houdini}
C.~Flanagan and K.~R.~M. Leino, ``{Houdini, an Annotation Assistant for
  ESC/Java},'' in \emph{International Symposium of Formal Methods Europe on
  Formal Methods for Increasing Software Productivity (FME)}.\hskip 1em plus
  0.5em minus 0.4em\relax Springer-Verlag, 2001, p. 500–517.

\bibitem{daikon2007}
M.~D. Ernst, J.~H. Perkins, P.~J. Guo, S.~McCamant, C.~Pacheco, M.~S. Tschantz,
  and C.~Xiao, ``{The Daikon System for Dynamic Detection of Likely
  Invariants},'' \emph{Science of computer programming}, vol.~69, no. 1–3, p.
  35–45, 2007.

\bibitem{LMN14}
P.~Garg, C.~L{"{o}}ding, P.~Madhusudan, and D.~Neider, ``{ICE:} {A} robust
  framework for learning invariants,'' in \emph{Computer Aided Verification
  ({CAV})}.\hskip 1em plus 0.5em minus 0.4em\relax Springer, 2014.

\bibitem{Garg2016}
P.~Garg, D.~Neider, P.~Madhusudan, and D.~Roth, ``Learning invariants using
  decision trees and implication counterexamples,'' \emph{SIGPLAN Not.},
  vol.~51, no.~1, p. 499–512, 2016.

\bibitem{DBLP:conf/memocode/LiDS11}
W.~Li, L.~Dworkin, and S.~A. Seshia, ``Mining assumptions for synthesis,'' in
  \emph{International Conference on Formal Methods and Models}.\hskip 1em plus
  0.5em minus 0.4em\relax {IEEE}, 2011, pp. 43--50.

\bibitem{DBLP:conf/fmcad/AlurMT13}
R.~Alur, S.~Moarref, and U.~Topcu, ``Counter-strategy guided refinement of
  {GR(1)} temporal logic specifications,'' in \emph{Formal Methods in
  Computer-Aided Design, {FMCAD}}.\hskip 1em plus 0.5em minus 0.4em\relax
  {IEEE}, 2013, pp. 26--33.

\bibitem{keegan2020control}
M.~Keegan, V.~A. Braberman, N.~D'Ippolito, N.~Piterman, and S.~Uchitel,
  ``Control and discovery of environment behaviour,'' \emph{IEEE Transactions
  on Software Engineering}, 2020.

\bibitem{sato2020constrained}
S.~Sato, M.~Waga, and I.~Hasuo, ``Constrained optimization for falsification
  and conjunctive synthesis,'' \emph{arXiv preprint arXiv:2012.00319}, 2020.

\bibitem{DBLP:journals/tcad/JinDDS15}
X.~Jin, A.~Donz{\'{e}}, J.~V. Deshmukh, and S.~A. Seshia, ``Mining requirements
  from closed-loop control models,'' \emph{IEEE Transactions on Computer-Aided
  Design of Integrated Circuits and Systems}, vol.~34, no.~11, pp. 1704--1717,
  2015.

\bibitem{DBLP:conf/hybrid/JinDDS13}
X.~Jin, A.~Donz{\'{e}}, J.~V.~Deshmukh, and S.~A.~Seshia, ``Mining requirements
  from closed-loop control models,'' in \emph{International conference on
  Hybrid systems: computation and control, {HSCC}}.\hskip 1em plus 0.5em minus
  0.4em\relax {ACM}, 2013.

\bibitem{6032536}
R.~V. {Borges}, A.~{d'Avila Garcez}, L.~C. {Lamb}, and B.~{Nuseibeh},
  ``Learning to adapt requirements specifications of evolving systems,'' in
  \emph{International Conference on Software Engineering (ICSE)}.\hskip 1em
  plus 0.5em minus 0.4em\relax IEEE, 2011.

\bibitem{DBLP:conf/asplos/Solar-LezamaTBSS06}
A.~Solar{-}Lezama, L.~Tancau, R.~Bod{\'{\i}}k, S.~A. Seshia, and V.~A.
  Saraswat, ``Combinatorial sketching for finite programs,'' in
  \emph{International Conference on Architectural Support for Programming
  Languages and Operating Systems ({ASPLOS})}.\hskip 1em plus 0.5em minus
  0.4em\relax {ACM}, 2006.

\end{thebibliography}

\begin{IEEEbiography}[{\includegraphics[width=1in,height=1.25in,clip,keepaspectratio]{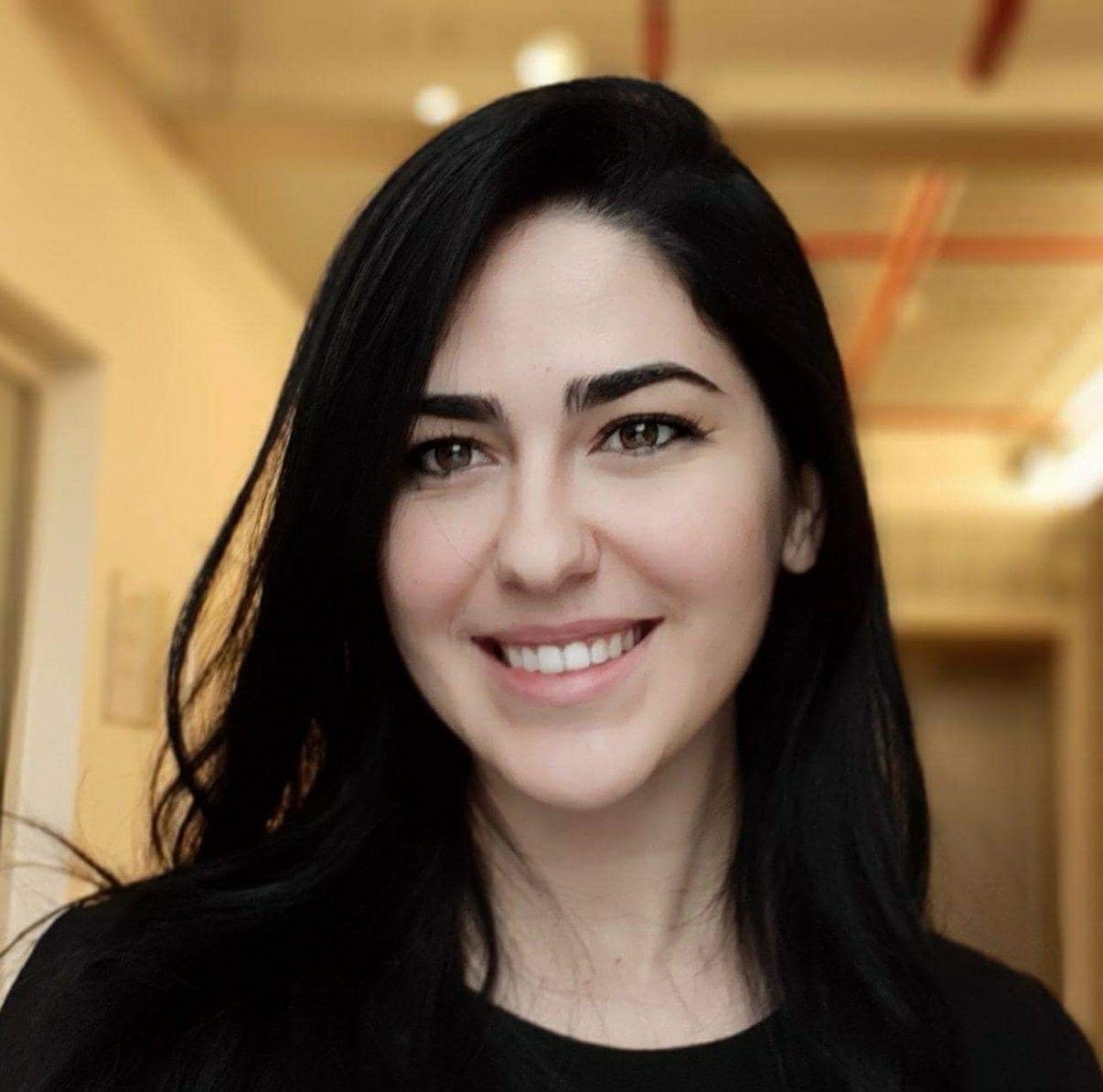}}]{Khouloud Gaaloul}
is a Ph.D. candidate at the Inter-disciplinary Centre for Security, Reliability and Trust(SnT), University of Luxembourg. She received her Engineering degree from the Higher National School of Engineering of Tunis (ENSIT) in 2016. Her research interests include software quality, search-based software engineering and testing and applied machine learning.
\end{IEEEbiography}

\begin{IEEEbiography}[{\includegraphics[width=1in,height=1.25in,clip,keepaspectratio]{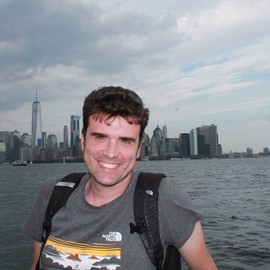}}]{Claudio Menghi}
is a Research Associate at the  Interdisciplinary Centre for Security, Reliability and Trust (SnT), at the University of Luxembourg. After receiving his Ph.D. at Politecnico di Milano in 2015, he was a post-doctoral researcher at Chalmers University of Technology and University of Gothenburg. His research interests are in software engineering, with a special interest in cyber-physical systems (CPS) and formal verification.
\end{IEEEbiography}

\begin{IEEEbiography}[{\includegraphics[width=1in,height=1.25in,clip,keepaspectratio]{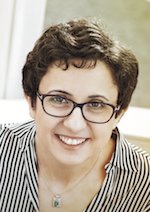}}]{Shiva Nejati}
is an Associate Professor at the School of Electrical Engineering and Computer Science of the University of Ottawa and a part-time Research Scientist at the SnT Centre for Security, Reliability, and Trust, University of Luxembourg. From 2009 to 2012, she was a researcher at Simula Research Laboratory in Norway. She received her Ph.D. degree from the University of Toronto, Canada in 2008. Nejati’s research interests are in software engineering, focusing on model-based development, software testing, analysis of cyber-physical systems, search-based software engineering and formal and empirical software engineering methods. Nejati has coauthored over 50 journal and conference papers, and regularly serves on the program committees of international conferences in the area of software engineering. She has for the past ten years been conducting her research in close collaboration with industry partners in telecommunication, maritime, energy, automotive and aerospace sectors.
\end{IEEEbiography}

\begin{IEEEbiography}[{\includegraphics[width=1in,height=1.25in,clip,keepaspectratio]{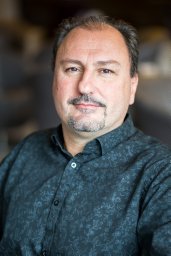}}]{Lionel C. Briand}
is professor of software engineering and has shared appointments between (1) The University of Ottawa, Canada and (2) The SnT Centre for Security, Reliability, and Trust, University of Luxembourg. He is currently running multiple collaborative research projects with companies in the automotive, satellite, financial, and legal domains. Lionel has held various engineering, academic, and leading positions in six countries. He was one of the founders of the ICST conference (IEEE Int. Conf. on Software Testing,Verification, and Validation, a CORE A event) and its first general chair. He was also EiC of Empirical Software Engineering (Springer) for 13 years and led, in collaboration with first Victor Basili and then Tom Zimmermann, the journal to the top tier of the very best publication venues in software engineering. Lionel was elevated to the grade of IEEE Fellow in 2010 for his work on testing object-oriented systems. He was granted the IEEE Computer Society Harlan Mills award and the IEEE Reliability Society engineer-of-the-year award for his work on model-based verification and testing, respectively in 2012 and2013. He received an ERC Advanced grant in 2016 – on the topic of modeling and testing cyber-physical systems – which is the most prestigious individual research award in the European Union. Most recently, he was awarded a Canada Research Chair (Tier 1) on ‘‘Intelligent Software Dependability and Compliance’’. His research interests include: software testing and verification, model-driven software development, applications of AI in software engineering, and empirical software engineering.
\end{IEEEbiography}

\begin{IEEEbiography}[{\includegraphics[width=1in,height=1.25in,clip,keepaspectratio]{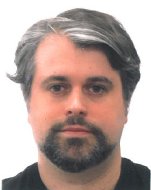}}]{Yago Isasi Parache} is Head of the Software division at LuxSpace Sàrl, and leads the development of satellite simulators, satellite on-board software, satellite data applications, software product assurance, and software tooling for space software development. He holds a Diploma of Advanced Studies (DEA - Master Degree) in Applied Mathematics from Universitat Politècnica de Catalunya, a Diploma of Advanced Studies (DEA - Master Degree) in Astrophysics, Particle Physics, and Cosmology from Universita de Barcelona, and a degree in Industrial Engineering from Universitat Politècnica de Catalunya.
\end{IEEEbiography}

\end{document}